\documentclass[]{aa}
\usepackage{graphicx}
\usepackage{txfonts}

\def\gsimeq
{\hbox{\raise0.5ex\hbox{$>\lower1.06ex\hbox{$\kern-1.07em{\sim}$}$}}}
\def\lsimeq
{\hbox{\raise0.5ex\hbox{$<\lower1.06ex\hbox{$\kern-1.07em{\sim}$}$}}}

\def\aj{{AJ}}

\def\bet{$_{\beta}$}

\def\apj{{ApJ}}
\def\apjs{{ApJS}}

\def\deg{$^{\circ}$}

\def\lamb{$\lambda$}

\def\micron{{$\mu$m}}
\def\mnras{{MNRAS}}
\def\nat{{Nature}}

\def\percm2{cm$^{-2}$}

\def\oiii{[OIII]$_{\lambda5007}$}
\def\ha{H$_{\alpha}$}

\def\oi{[$\ion{O}{I}$]}
\def\nii{[$\ion{N}{II}$]}

\def\sii{[$\ion{S}{II}$]}

\def\cha{{\it Chandra}~}
\def\xmm{{\it XMM-Newton}~}

\def\ltsima{$\; \buildrel < \over \sim \;$}
\def\simlt{\lower.5ex\hbox{\ltsima}}
\def\gtsima{$\; \buildrel > \over \sim \;$}
\def\simgt{\lower.5ex\hbox{\gtsima}}

\def\lbol{L$_{bol}$} 
\def\hl{10$^{44}$ erg/s}

\def\hard{$_{2-10 keV}$}
\begin{document}
   \title{On the X-ray, optical emission 
   line and black hole mass properties of local Seyfert galaxies}


   \author{F. Panessa\inst{1}, L. Bassani\inst{2}, M. Cappi\inst{2},
   	M. Dadina\inst{2}, X. Barcons\inst{1}, F. J. Carrera\inst{1}, 
	L.C. Ho\inst{3}, K. Iwasawa\inst{4}    
          }

   \offprints{Francesca Panessa\\ \email{panessa@ifca.unican.es}}

   \institute{Instituto de F{\'\i}sica de Cantabria (CSIC-UC),
   Avda. de los Castros, 39005 Santander, Spain
             \and INAF-IASF, Via P. Gobetti 101, 40129 Bologna, Italy
	     \and The Observatories of the Carnegie Institution of Washington, 813 Santa Barbara St.
	     Pasadena, CA 91101
	     \and Max Planck Institut f\"{u}r Extraterrestrische Physik (MPE), Giessenbachstrasse 1,
	     D-85748 Garching, Germany   
}

\authorrunning{F. Panessa et al.}

   \abstract{
We investigate the relation between X-ray nuclear emission, 
optical emission line luminosities and black hole masses
for a sample of 47 Seyfert galaxies.
The sample, which has been selected from the Palomar optical
spectroscopic survey of nearby galaxies (Ho, Filippenko \& Sargent 1997),
covers a wide range of nuclear powers, from L$_{2-10 keV}$ $\sim$ 10$^{43}$ erg/s 
down to very low luminosities
(L$_{2-10 keV}$ $\sim$ 10$^{38}$ erg/s). 
Best available data from Chandra, XMM-Newton
and, in a few cases, ASCA observations have been considered. Thanks to the good spatial resolution 
available from these observations and 
a proper modeling of the various spectral components, it has been
possible to obtain accurate nuclear X-ray luminosities
not contaminated by off-nuclear sources and/or diffuse emission.
X-ray luminosities have then been corrected taking into 
account the likely candidate Compton thick sources, which are
a high fraction ($>$ 30\%) among type 2 Seyferts in our sample.
The main result of this study is that
we confirm strong linear correlations between 2-10 keV, [OIII]$\lambda$5007,
H$_{\alpha}$ luminosities 
which show the same slope as quasars and luminous Seyfert galaxies, 
independent of the level of nuclear activity
displayed. Moreover, despite the wide range
of Eddington ratios (L/L$_{Edd}$) tested here (six orders
of magnitude, from 0.1 down to $\sim$ 10$^{-7}$), 
no correlation is found between
the X-ray or optical emission line luminosities and the black hole mass.
Our results suggest that Seyfert nuclei in our sample 
are consistent with being a scaled-down version
of more luminous AGN.
   \keywords{accretion, accretion disks - X-rays: galaxies - galaxies: Seyfert - galaxies: nuclei}
   }

   \maketitle
%

\section{Introduction}

The fundamental paradigm on which our understanding of Active Galactic 
Nuclei (AGN) activity is
based is that the accretion of matter onto a 
super massive black hole (SMBH) powers the energy
emission of these objects. Starting from this paradigm,
the unification scheme (Unified Model, UM) proposed by 
Antonucci \& Miller (1985)
relocates the various families of AGN 
within a single scenario in which
the orientation of the emitter-absorber-observer system is responsible
for the different spectral properties exhibited. In this scenario,
type 2 objects are the ones in which the absorber (dusty torus)
intercepts the line of sight, otherwise our view is directed into the
nucleus and we are dealing with type 1 objects. 
Within the UM, the only other free parameter is, besides the orientation, 
the mechanism responsible for triggering strong radio emission
and associated jets which are present in 
$\sim$ 10\% of the AGN population.

In recent years, the number of cases in which 
the zeroth-order UM predictions seem not to be completely adequate to explain
all observational evidences 
is increasing, both in the local and in the distant universe. 
Type 1 AGN with significant absorption have been found 
(Cappi et al. 2006, hereafter C06, Mateos et al. 2005, Fiore et al. 2001a, etc.)
as well as type 2 AGN without X-ray absorption
(Caccianiga et al. 2004, Corral et al. 2005, Barcons, 
Carrera \& Ceballos 2003, Panessa \& Bassani 2002,
Pappa et al. 2001). In particular, it is not clear whether at very low luminosities 
the predictions of UM are still valid (Panessa \& Bassani 2002, Ho et al. 2001).
The key for the comprehension of the whole AGN phenomenon
seems to reside in a combination of the UM hypothesis and
the fundamental parameters of AGN, such as black hole mass, 
Eddington ratio, and perhaps the black hole spin.
After the discovery that SMBHs
reside at the center 
of most, if not all, galaxies in the nearby universe
(Kormendy \& Richstone 1995, Magorrian et al. 1998) and that
a large fraction of them are active (at least 40\%,
Ho, Filippenko \& Sargent 1997b), it is of fundamental
importance to understand the accretion physics in AGN
and what triggers the different levels of activity.

One of the distinctive characteristic of nearby nuclei
is their intrinsic faintness, i.e. L$_{Bol}$ $<$ 10$^{44}$ erg/s
(Ho 2003), as well as their low level of activity;
in terms of Eddington luminosity most of them
have L/L$_{Edd}$ $<$ 10$^{-2}$ compared to
L/L$_{Edd}$ $\sim$ 1 of luminous AGN.
Whether low luminosity AGN (LLAGN) are a scaled-down luminosity version
of classical AGN or objects powered by different
physical mechanism is a debated issue. 
It is not clear in fact, whether
LLAGN are powered by radiatively inefficient accretion flows,
such as advection dominated accretion flows (ADAF)
and their variants (Narayan \& Yi 1994,
Abramowicz 1997) instead of
the standard geometrically thin optically thick accretion disk
typically proposed as the accretion mechanism acting
in the central regions of luminous AGN (Shakura \& Sunyaev 1973). 
LLAGN could also represent scaled up versions of black
hole binaries in the steady-jet, hard X-ray state,
as pointed out by the scaling relations reported
in Merloni et al. (2003) and Falcke et al. (2004).
On one hand, ADAF models are able to predict some of 
the spectral properties observed in many LLAGN,
such as the lack of the 'big blue bump' (Ho 1999).
On the other hand, some LLAGN show properties
which are common to luminous AGN, such as
the observed correlations between
optical emission lines and ionizing continuum
(Ho \& Peng 2001) or X-ray emission 
(Ward et al. 1988, Terashima et al. 2000, Ho et al. 2001).

X-rays are one of the most direct
evidences of nuclear activity and are, therefore, fundamental
to study the accretion processes.
The new generation X-ray telescopes Chandra and
XMM-Newton provide a much better spatial and spectral resolution
than previous satellites.
The high spatial resolution allows the 
detection of genuine low-luminosity AGN
and the separation of the contribution
of enhanced circumnuclear star formation 
from bona-fide AGN with compact nuclear
sources (Ho et al. 2001).
The spectral resolution allows a good characterization
of the X-ray spectral features. 
It is particularly important to determine the intrinsic absorbing
column density accurately, since heavy obscuration suppresses the soft
X-ray emission by a large factor and alters the spectral shape. As shown
by the study of the distance-limited sample of nearby Seyfert galaxies
(C06), the X-ray luminosity distribution is strongly affected by a
non-negligible number of Compton-thick sources
(N$_{H}$ $>$ 10$^{24}$ cm$^{-2}$), which cannot be assessed
by the data below 10 keV alone.

Closely related to the theoretical and observational
issues in LLAGN is the determination of the mass
of the 'massive dark objects', presumably black holes,
located at the center of nearby galaxies, via stellar/gas kinematics (Gebhardt et al. 2003),
via black hole mass (M$_{BH}$) and stellar velocity dispersion 
(Tremaine et al. 2002) or galaxy bulge mass (Richstone et al. 1998)
correlations. The observed radiative output (e.g., X-ray luminosities)
combined with M$_{BH}$ estimates, allows us to measure the
Eddington ratios and, therefore investigate the
fundamental scaling of black hole properties
with M$_{BH}$ and accretion rate, $\dot{m}$.

We have chosen to investigate LLAGN
and their relation with luminous AGN by studying the properties
of a well defined sample of nearby Seyfert galaxies  
selected from Ho, Filippenko \& Sargent (1997a), hereafter HFS97. 
In this paper, we focus on the X-ray, H$_{\alpha}$ and [OIII] 
emission line properties. The
estimates of the central BH masses are then used to test the activity
levels of the sources. The strength of our approach resides in the capability
to trace the absorption both via X-ray spectra and model
independent diagnostics. 
For our analysis, we took advantage of the
results presented in a companion paper (C06), 
where the XMM-Newton data were presented for a distance limited sub-sample of the
HFS97 Seyfert list.   

The paper is organized as follows:
a description of the sample is given in Sec. 2, details of the X-ray observations 
and data reduction can be found in Sec. 3, in Sec. 4 
we discuss the fraction of heavily absorbed sources; relations between 
X-ray/optical emission line luminosities with black hole masses are discussed in Sec. 6.
The results and the effects of the incompleteness of the sample are discussed in Sec. 7.
Finally, conclusions are summarized in Sec. 8. A description of the 
X-ray spectra analyzed in this work is deferred to the Appendix.

\section{The sample}

Our sample of Seyfert galaxies has been derived from the Palomar 
optical spectroscopic survey of nearby galaxies (Ho, Filippenko, \& Sargent 
1995). From this survey, high-quality optical spectra of 486 bright 
($B_T\,\leq$ 12.5 mag), northern ($\delta\,>$ 0\deg) galaxies 
have been taken and a comprehensive, homogeneous catalog of spectral 
classifications of all galaxies have been obtained (HFS97). 
The Palomar survey is complete to $B_T$ = 12.0 mag and 80\% complete 
to $B_T$ = 12.5 mag (Sandage, Tammann, \& Yahil 1979).  
For the purpose of our study
this is one of the best samples available up to now. In fact,
it offers an accurate optical classification and 
the opportunity of detecting weak nuclei. 
Finally, the sample covers a large range of AGN
luminosities (\lbol~$\sim$~10$^{41}$-\hl) 
making it ideal for exploring possible trends with AGN power.

Here, the spectroscopic classification system of the Palomar survey
is briefly summarized
(see HFS97 for a more accurate description). 
The relative strength of the low-ionization optical forbidden lines (\oi\ 
\lamb\lamb 6300, 6364, \nii\ \lamb\lamb 6548, 6583, \sii\ \lamb\lamb 6716, 
6731) compared to the hydrogen Balmer lines determines the 
classification of emission-line nuclei into two classes: H II nuclei 
(powered by stars) and AGN (powered by black-hole accretion).
The separation between LINERs and Seyferts is instead given
by the ratio \oiii/H\bet  which corresponds 
to the ionization state of the narrow-line gas in AGN,
i.e. \oiii/H\bet\ $<$ 3 for LINERs and \oiii/H\bet\ $\geq$ 3 for Seyferts. 
Emission line nuclei having [OI] strengths
intermediate between those of H II nuclei and LINERs
are classified as "transition objects". 
Symbols used are: L = LINER, 
T = ``transition object'' (LINER + HII nucleus), and S =  Seyfert. 
The classification in ``type~1'' or ``type~2''
depends on the presence or absence of broad 
permitted lines. The measurement of the 
relative strength of the broad component of the 
hydrogen Balmer lines lead to 
subdivisions in the classification 
(type 1.0, 1.2, 1.5, 1.8 and 1.9; see Osterbrock 1981).

From the entire HFS97 sample we have extracted all
Seyfert galaxies. The total sample of 60 Seyfert galaxies 
includes 39 type~2 
(type 2 and 1.9) and 13 type~1 AGN (type 1.0, 1.2, 1.5).
Eight objects, which are placed near the boundary between
Seyfert and LINER, HII or transition classification, with
a double classification (e.g., S2/T2, L2/S2, H/S2, etc.),
have been included in the final sample. Hereafter we refer to these objects
as 'mixed Seyferts'.

Seyfert galaxies classified as type 2 and 1.9 have been grouped
into a more general 'type 2' classification,  while type 1.0, 1.2 and 1.5
have been grouped in the 'type 1' class.
Type 2 and type 1.9 sources are normally both absorbed objects,
while the type 1 group is referred to objects which
are normally not affected by heavy absorption.

Two sources of the sample, NGC 4395 and NGC 4579,
which have been classified by HFS97 as S1.8 and S1.9/L1.9 respectively,
have been reclassified as type 1.5. 
A broad component is present in a number of optical 
(Filippenko \& Sargent  1989) and ultraviolet 
(Filippenko, Ho \& Sargent 1993) emission lines of NGC 4395.
Extremely broad permitted lines have been detected in NGC 4579.
{\it HST} observations have revealed an H$_{\alpha}$
component with FWZI of $\sim$ 18000 km/s  
(Barth et al. 1996 and Barth et al. 2001).

In objects like NGC 3608, NGC 3941, NGC 4472 and NGC 6482, 
the difficulty in the starlight subtraction process has 
lead to uncertainties in the classification (HFS97).
Finally, the classification of NGC 185 is also uncertain, i.e.,
it is a dwarf spheroidal galaxy
whose Seyfert-like line ratios maybe produced by 
stellar processes (Ho \& Ulvestad 2001).

Table~1 lists the properties of the host galaxies of the sample.
Data for Cols. (2)--(8) are taken from the
compilation of Ho \& Ulvestad (2001) and references therein.  
Distances for a few objects have been updated with more recent
estimates (references are indicated in Col. (5)).
The median distance of the sample galaxies is 25.7$\pm$ 17.7 Mpc.
The nearest galaxy is NGC 185 (one of the
companions of M 31) at D = 0.64 Mpc and the farthest
is NGC 5548 at D = 70.2 Mpc, so we are sampling the local universe.

\begin{table*}
\scriptsize{
\label{tab=sample}
\begin{center}
\caption{\bf The Seyfert galaxy sample}
\begin{tabular}{lccccccc}
\hline
\hline
\multicolumn{1}{c}{Name} &
\multicolumn{1}{c}{Seyfert} &
\multicolumn{1}{c}{Hubble} &
\multicolumn{1}{c}{B} &
\multicolumn{1}{c}{Distance} &
\multicolumn{1}{c}{Ref.}&
\multicolumn{1}{c}{R.A.} &
\multicolumn{1}{c}{Dec.} \\
\multicolumn{1}{c}{} &
\multicolumn{1}{c}{Type} &
\multicolumn{1}{c}{Type} &
\multicolumn{1}{c}{(mag)} &
\multicolumn{1}{c}{(Mpc)} &
\multicolumn{1}{c}{} &
\multicolumn{1}{c}{(2000)} &
\multicolumn{1}{c}{(2000)} \\
\multicolumn{1}{c}{(1)} &
\multicolumn{1}{c}{(2)} &
\multicolumn{1}{c}{(3)} &
\multicolumn{1}{c}{(4)} &
\multicolumn{1}{c}{(5)}&
\multicolumn{1}{c}{(6)} &
\multicolumn{1}{c}{(7)} &
\multicolumn{1}{c}{(8)} \\
\hline
\hline
NGC185  & S2      & dE3pec     &10.10 & 0.64 & a&   00 38 57.4  &  +48 20 14.4 \\
NGC676  & S2:     & S0/a:spin  &10.50 & 19.5 & b&   01 48 57.4  &  +05 54 25.7 \\ 
NGC777  & S2/L2   & E1         &12.49 & 66.5 & b&   02 00 15.0  &  +31 25 45.5 \\
NGC1058 & S2      & Sc         &11.82 &  9.1 & b&   02 43 30.2  &  +37 20 27.2 \\
NGC1068 & S1.9    & Sb         & 9.61 & 14.4 & b&   02 42 40.7  &  -00 00 47.6 \\
NGC1167 & S2      & S0-        &13.38 & 65.3 & b&   03 01 42.4  &  +35 12 20.7 \\
NGC1275 & S1.5    & Pec        &12.64 & 70.1 & b&   03 19 48.2  &  +41 30 42.4 \\
NGC1358 & S2      & SAB0/a     &13.04 & 53.6 & b&   03 33 39.7  &  -05 05 21.8 \\
NGC1667 & S2      & SABc       &12.77 & 61.2 & b&   04 48 37.1  &  -06 19 11.9 \\
NGC2273 & S2      & SBa        &12.55 & 28.4 & b&   06 50 08.7  &  +60 50 45.0 \\
NGC2336 & L2/S2   & SABbc      &10.61 & 33.9 & b&   07 27 03.8  &  +80 10 39.6 \\
NGC2639 & S1.9    & Sa?        &12.46 & 42.6 & b&   08 43 38.0  &  +50 12 20.3 \\
NGC2655 & S2      & SAB0/a     &10.96 & 24.4 & b&   08 55 38.8  &  +78 13 25.2 \\    
NGC2683 & L2/S2   & SAb        & 9.62 &  7.7 & a&   08 52 41.7  &  +33 25 10.4 \\
NGC2685 & S2/T2:  & SB0+ pec   &12.12 & 16.2 & b&   08 55 34.8  &  +58 44 01.6 \\  
NGC3031 & S1.5    & Sab        & 7.89 &  3.5 & c&   09 55 33.2  &  +69 03 55.0 \\ 
NGC3079 & S2      & SBc spin   &11.54 & 17.3 & d&   10 01 58.5  &  +55 40 50.1 \\  
NGC3147 & S2      & Sbc        &11.43 & 40.9 & b&   10 16 53.3  &  +73 24 02.4 \\
NGC3185 & S2:     & SB0/a      &12.99 & 21.3 & b&   10 17 38.7  &  +21 41 17.2 \\ 
NGC3227 & S1.5    & SABa pec   &11.10 & 20.6 & b&   10 23 30.6  &  +19 51 53.9 \\  
NGC3254 & S2      & Sbc        &12.41 & 23.6 & b&   10 29 19.9  &  +29 29 29.6 \\
NGC3486 & S2      & SABc       &11.05 &  7.4 & b&   11 00 24.1  &  +28 58 31.6 \\   
NGC3489 & T2/S2   & SAB+       &11.15 &  6.4 & b&   11 00 18.6  &  +13 54 04.0 \\
NGC3516 & S1.2    & SB0        &12.50 & 35.7 & b&   11 06 47.5  &  +72 34 06.9 \\
NGC3608 & L2/S2   & E2         &11.69 & 23.4 & b&   11 16 59.1  &  +18 08 54.6 \\
NGC3627 & T2/S2   & SABb       & 9.13 &  6.6 & b&   11 20 15.1  &  +12 59 21.6 \\
NGC3655 & H/S2    & SAc        &12.08 & 26.5 & b&   11 22 54.7  &  +16 35 22.0 \\
NGC3735 & S2      & Sc: spin   &12.50 & 41.0 & b&   11 35 57.5  &  +70 32 07.7 \\
NGC3941 & S2:     & SB0        &11.25 & 12.2 & a&   11 52 55.4  &  +36 59 10.5 \\  
NGC3976 & S2      & SABb       &12.30 & 37.7 & b&   11 55 57.3  &  +06 44 57.0 \\
NGC3982 & S1.9    & SABb:      &11.78 & 20.5 & e&   11 56 28.1  &  +55 07 30.6 \\  
NGC4051 & S1.2    & SABbc      &10.83 & 17.0 & b&   12 03 09.6  &  +44 31 52.8 \\  
NGC4138 & S1.9    & S0+        &12.16 & 13.8 & a&   12 09 29.9  &  +43 41 06.0 \\  
NGC4151 & S1.5    & SABab:     &11.50 & 20.3 & b&   12 10 32.6  &  +39 24 20.6 \\  
NGC4168 & S1.9:   & E2         &12.11 & 31.7 & f&   12 12 17.3  &  +13 12 17.9 \\  
NGC4169 & S2      & S0         &13.15 & 50.4 & b&   12 12 18.9  &  +29 10 44.0 \\
NGC4235 & S1.2    & Sa spin    &12.62 & 35.1 & b&   12 17 09.9  &  +07 11 29.1 \\
NGC4258 & S1.9    & SABbc      & 9.10 &  7.2 & a&   12 18 57.5  &  +47 18 14.3 \\   
NGC4378 & S2      & Sa         &12.63 & 35.1 & b&   12 25 18.1  &  +04 55 31.6 \\
NGC4388 & S1.9    & Sb: spin   &11.76 & 16.7 & b&   12 25 46.7  &  +12 39 40.9 \\  
NGC4395 & S1*     & Sm:        &10.64 &  2.6 & b&   12 25 48.9  &  +33 32 47.8 \\   
NGC4472 & S2::    & E2         & 9.37 & 16.7 & b&   12 29 46.8  &  +07 59 59.9 \\  
NGC4477 & S2      & SB0?       &11.38 & 16.8 & b&   12 30 02.2  &  +13 38 11.3 \\  
NGC4501 & S1.9    & Sb         &10.36 & 16.8 & b&   12 31 59.3  &  +14 25 13.4 \\ 
NGC4565 & S1.9    & Sb? spin   &10.42 &  9.7 & a&   12 36 21.1  &  +25 59 13.5 \\   
NGC4579 & S1*     & SABb       &10.48 & 16.8 & b&   12 37 43.5  &  +11 49 04.9\\  
NGC4639 & S1.0    & SABbc      &12.24 & 22.9 & b&   12 42 52.5  &  +13 15 24.1 \\  
NGC4698 & S2      & Sab        &11.46 & 16.8 & b&   12 48 22.9  &  +08 29 14.8 \\  
NGC4725 & S2:     & SABab pec  &10.11 & 13.2 & a&   12 50 26.7  &  +25 30 02.3 \\  
NGC5033 & S1.5    & Sc         &10.75 & 18.7 & b&   13 13 27.5  &  +36 35 37.8 \\  
NGC5194 & S2      & Sbc pec    & 8.96 &  8.4 & b&   13 29 52.4  &  +47 11 40.8 \\   
NGC5273 & S1.5    & S0         &12.44 & 16.5 & a&   13 42 08.3  &  +35 39 15.2 \\  
NGC5395 & S2/L2   & Sb pec     &12.10 & 46.7 & b&   13 58 38.0  &  +37 25 28.3 \\
NGC5548 & S1.5    & S0/a       &13.30 & 70.2 & b&   14 17 59.5  &  +25 08 12.4 \\
NGC5631 & S2/L2   & S0         &12.41 & 27.8 & a&   14 26 33.3  &  +56 34 58.3 \\
NGC6482 & T2/S2   & E          &11.84 & 52.3 & b&   17 51 48.9  &  +23 04 19.1 \\
NGC6503 & T2/S2   & SAcd       &10.11 & 5.2  & g&   17 49 26.6  &  +70 08 40.1 \\
NGC6951 & S2      & SABbc      &11.64 & 24.1 & b&   20 37 14.4  &  +66 06 19.7 \\
NGC7479 & S1.9    & SBc        &11.60 & 32.4 & b&   23 04 56.7  &  +12 19 23.2 \\
NGC7743 & S2      & SB0+       &12.38 & 24.4 & b&   23 44 21.4  &  +09 56 03.6 \\
\hline
\hline
\end{tabular}
\end{center}
Notes: Col. (1): Galaxy name; Col. (2): optical classification, '*'= objects 
with a changed classification with respect to the original given by 
HFS97. The quality rating is given by ``:'' 
and ``::'' for uncertain and highly uncertain classification, respectively,
as reported in HFS97.; Col. (3): Hubble type; 
Col. (4) total apparent $B$ magnitude of the galaxy; 
Col. (5): galaxy distances; Col. (6): 
references on distances: (a) Tonry et al. (2001); (b) Tully (1988);  
(c) Paturel et al. (2002); (d) Cecil et al. 2002; (e) Stetson et al. (2001); 
(f) Merritt \& Ferrarese (2001); (g) Karachentsev \& Sharina (1997); 
Col. (7)-(8) optical position in epoch J2000.
}
\end{table*}

\section{The X-ray data}

An homogeneous and standard X-ray data analysis has been carried out on
our selected Seyfert sample using \cha and \xmm observations. 
\cha and \xmm observations are available for 39
objects of the sample with 22 objects having 
observations with both observatories. 
Most of the XMM-Newton observations are derived 
from an EPIC Guaranteed Time survey of a distance-limited 
sample of 27 Seyfert galaxies which
have been published in a companion paper
(C06) and we refer to that work
for details.

To complement the X-ray information on 
the whole sample, a search in the literature 
for observations with previous X-ray observatories
(operating in the 2-10 keV energy range) has also been carried out. 
{\it ASCA} observations have been found 
for 8 further objects, references for those data taken from the
literature are given in Table~$\ref{x}$, except for NGC 3982 and NGC 4235 
(briefly discussed in the Appendix) 
for which {\it ASCA} fluxes have been derived in this work. 
Adding all these data, 47 sources out of 60 objects have X-ray data available.

We used the CIAO 3.0 software for the 
Chandra data analysis to perform the data processing and 
calibration\footnote{All the data processing have been carried 
out following analysis threads 
on the Chandra web site: http://cxc.harvard.edu/ciao/threads/index.html}. 
Starting from level 1 files, new level 2 event files were generated using the
latest calibrations.
We applied the pixel randomization introduced by the CXC (Chandra X-ray
observatory Center) standard data 
processing (SDP) to avoid the instrumental "gridded" appearance of the data 
and any possible aliasing effects associated with this spatial grid.
Finally, we examined light curves in order to clean the datasets 
for periods of anomalous background rates.

Most observations have been taken in the standard mode 
that allows a read-out mode of the full chip
every 3.2s. For many bright sources in the sample more than one observation 
is often available. In this case, we have chosen the data set without gratings and, 
in order to minimize pile-up effects, the data set with 1/8 or 1/2 chip sub-array
mode.

\begin{table*}[!htb]
\begin{center}
\caption{\bf Multi-wavelength luminosities and M$_{BH}$ of the total X-ray sample}
\footnotesize{
\label{x}
\begin{tabular}{cccrrccrrrrr}
\hline
\hline
\multicolumn{1}{c}{Name} &
\multicolumn{1}{c}{Class}&
\multicolumn{1}{c}{Sat} &
\multicolumn{1}{r}{Log F$_{2-10 keV}$}&
\multicolumn{1}{r}{Log L$_{2-10 keV}$}&
\multicolumn{1}{r}{Ref.}&
\multicolumn{1}{r}{CThick}&
\multicolumn{1}{r}{Log L$_{[OIII]}$}&
\multicolumn{1}{r}{Log L$_{H\alpha}$}&
\multicolumn{1}{r}{Log M$_{BH}$}&
\multicolumn{1}{c}{Method} &
\multicolumn{1}{r}{Ref.}\\
\multicolumn{1}{c}{(1)} &
\multicolumn{1}{c}{(2)} &
\multicolumn{1}{c}{(3)} &
\multicolumn{1}{c}{(4)} &
\multicolumn{1}{c}{(5)}&
\multicolumn{1}{c}{(6)} &
\multicolumn{1}{c}{(7)} &
\multicolumn{1}{c}{(8)} &
\multicolumn{1}{c}{(9)} &
\multicolumn{1}{c}{(10)} &
\multicolumn{1}{c}{(11)} &
\multicolumn{1}{c}{(12)} \\
\hline
NGC676  &S2:    &X& -13.62 & 40.79	& 1 & $\surd$	  &	39.04	 &     38.62	 &  -	   & -  	& - \\  
NGC1058 &S2     &C&$<$ -14.42 &$<$ 37.55	& 2 & ? 	  &	37.90	 &     38.16	 &  4.88   &I$_{\sigma}$& 8  \\     
NGC1068 &S1.9   &X& -11.31 & 42.84	& 1 & $\surd$	  &	41.91	 &     41.53	 &  7.20   & M  	& 1  \\     
NGC1167 &S2     &A& -13.39 & 42.07	& 3 & $\surd$	  &	40.76	 &     40.81	 &  -	   &  - 	& -  \\  
NGC1275 &S1.5   &X& -10.91 & 42.83		& 2 & $\times$    &	41.91	 &     42.07	 &  8.51   &I$_{\sigma}$& 2   \\ 
NGC1667 &S2     &A&  -13.1 & 42.31	& 3 & $\surd$	  &	40.42	 &     40.17	 &  7.88   &I$_{\sigma}$& 2   \\    
NGC2273 &S2     &A& -12.16 & 42.58	& 4 & $\surd$	  &	40.94	 &     40.92	 &  7.30   &I$_{\sigma}$& 2  \\  
NGC2639 &S1.9   &A& -12.49 & 40.82		& 5 & ? 	  &	40.40	 &     40.56	 &  8.02   &I$_{\sigma}$& 9  \\       
NGC2655 &S2     &A& -10.98 & 41.85		& 5 & $\times$    &	39.92	 &     40.03	 &  7.77   &I$_{\sigma}$& 9  \\     
NGC2683 &L2/S2  &C& -13.62 & 38.21		& 2 & ? 	  &	38.58	 &     37.69	 &  7.51   &I$_{\sigma}$& 9  \\  
NGC2685 &S2/T2: &X&:-12.53 & 39.94		& 1 & $\times$    &	38.96	 &     39.21	 &  7.15   &I$_{\sigma}$& 9   \\ 
NGC3031 &S1.5   &X& -10.89 & 40.25		& 1 & $\times$    &	38.56	 &     38.94	 &  7.80   & S  	& 1   \\
NGC3079 &S2     &C& -11.69 & 42.62	& 2 & $\surd$	  &	40.07	 &     40.89	 &  7.65   &I$_{\sigma}$& 3   \\ 
NGC3147 &S2     &C& -11.39 & 41.89		& 2 & $\times$    &	40.07	 &     40.01	 &  8.79   &I$_{\sigma}$& 3  \\  
NGC3185 &S2:    &X&  -13.7 & 40.79	& 1 & $\surd$	  &	39.90	 &     40.07	 &  6.06   &I$_{\sigma}$& 2   \\ 
NGC3227 &S1.5   &X& -10.94 & 41.74		& 1 & $\times$    &	40.51	 &     40.20	 &  7.59   & R  	& 4  \\   
NGC3486 &S2     &X& -12.93 & 38.86		& 1 & ? 	  &	37.99	 &     37.85	 &  6.14   &I$_{\sigma}$& 8  \\ 
NGC3489 &T2/S2  &C&$<$ -13.57 & $<$ 39.88	& 2 & $\surd$	  &	38.86	 &     38.48	 &  7.48   &I$_{\sigma}$& 5   \\
NGC3516 &S1.2   &C& -10.87 & 42.29		& 2 & $\times$    &	40.50	 &     39.91	 &  7.36   & R  	& 1    \\
NGC3608 &L2/S2  &C& -13.69 & 39.10		& 2 & $\times$    &	38.22	 &     38.23	 &  8.04   & S  	& 1   \\
NGC3627 &T2/S2  &C&$<$ -13.81 &$<$ 37.88	& 2 & ? 	  &	38.79	 &     39.35	 &  7.99   &I$_{\sigma}$& 5   \\ 
NGC3941 &S2:    &X& -13.35 & 38.88		& 1 & $\times$    &	38.63	 &     38.78	 &  8.15   &I$_{\sigma}$& 7   \\
NGC3982 &S1.9   &A& -13.28 & 41.18	& 2 & $\surd$	  &	40.50	 &     39.87	 &  6.09   &I$_{\sigma}$& 2  \\  
NGC4051 &S1.2   &C&  -11.2 & 41.31		& 2 & $\times$    &	39.81	 &     39.68	 &  6.11   & R  	& 4  \\ 
NGC4138 &S1.9   &X& -11.04 & 41.29		& 1 & $\times$    &	 38.74   &     38.69	 &  7.75   &I$_{\sigma}$& 9  \\ 
NGC4151 &S1.5   &X&  -10.2 & 42.47		& 1 & $\times$    &	41.47	 &     40.94	 &  7.18   & R  	& 4 \\   
NGC4168 &S1.9:  &X& -13.19 & 39.87		& 1 & $\times$    &	38.46	 &     38.93	 &  7.95   &I$_{\sigma}$& 6 \\    
NGC4235 &S1.2   &A& -10.92 & 42.22		& 2 & $\times$    &	 40.41   &     40.97	 &  -	   &	-	& -  \\    
NGC4258 &S1.9   &C& -10.91 & 40.86		& 2 & $\times$    &	39.07	 &     38.65	 &  7.61   & M  	& 1 \\     
NGC4388 &S1.9   &C& -10.78 & 41.72		& 2 & $\times$    &	40.54	 &     40.26	 &  6.80   &I$_{\sigma}$& 3   \\  
NGC4395 &S1     &X& -11.07 & 39.81		& 1 & $\times$    &	38.28	 &     38.53	 &  5.04  & S	       & 1   \\  
NGC4472 &S2::   &C&$<$ -13.18 &$<$ 39.32	& 2 & $\times$    & $<$ 37.81	 & $<$ 37.59	 &  8.80   &I$_{\sigma}$& 6   \\
NGC4477 &S2     &X& -12.85 & 39.65		& 1 & $\times$    &	39.04	 &     39.06	 &  7.92   &I$_{\sigma}$& 9  \\
NGC4501 &S1.9   &X& -12.91 & 39.59		& 1 & $\times$    &	39.23	 &     39.06	 &  7.90   &I$_{\sigma}$& 3  \\ 
NGC4565 &S1.9   &X&  -12.6 & 39.43		& 1 & $\times$    &	38.69	 &     38.72	 &  7.70   &I$_{\sigma}$& 3  \\ 
NGC4579 &S1     &C& -11.47 & 41.03		& 6 & $\times$    &	39.42	 &     39.72	 &  7.78   &I$_{\sigma}$& 3  \\  
NGC4639 &S1.0   &X& -12.55 & 40.22		& 1 & $\times$    &	38.71	 &     39.90	 &  6.85   &I$_{\sigma}$& 8 \\  
NGC4698 &S2     &X& -13.34 & 39.16		& 1 & $\times$    &	38.86	 &     38.74	 &  7.84   &I$_{\sigma}$& 9 \\  
NGC4725 &S2:    &X&  -13.4 & 38.89		& 1 & $\times$    &	38.58	 &     38.24	 &  7.49   &I$_{\sigma}$& 3 \\   
NGC5033 &S1.5   &X& -11.52 & 41.08		& 1 & $\times$    &	39.72	 &     40.42	 &  7.30   &I$_{\sigma}$& 3 \\  	      
NGC5194 &S2     &C& -12.77 & 40.91	& 2 & $\surd$	  &	39.90	 &     39.87	 &  6.95   &I$_{\sigma}$& 2	  \\	
NGC5273 &S1.5   &X& -11.13 & 41.36		& 1 & $\times$    &	39.03	 &     38.48	 &  6.51   &I$_{\sigma}$& 2	 \\  
NGC5548 &S1.5   &X&  -10.5 & 43.25		& 2 & $\times$    &	41.16	 &     40.26	 &  8.03   &I$_{\sigma}$& 4   \\									      
NGC6482 &T2/S2  &C& -13.33 & 40.16		& 2 & $\times$    &	-	 &     39.26	 &  8.75   &I$_{\sigma}$& 9\\	
NGC6503 &T2/S2  &C& -14.39 & 37.10		& 2 & ? 	  &	-	 &     37.41	 &  5.56   &I$_{\sigma}$& 8\\	
NGC7479 &S1.9   &X& -11.95 & 41.12		& 2 & ? 	  &	40.25	 &     40.67	 &  7.07   &I$_{\sigma}$& 9\\	
NGC7743 &S2     &A& -13.14 & 41.47	& 5 & $\surd$	  &	40.21	 &     40.24	 &  6.59   &I$_{\sigma}$& 2 \\  	   
\hline						    				    	  			    				     
\end{tabular}													    
\\					    				    	 				    
\scriptsize{Notes: Col. (1): galaxy name; Col. (2): optical classification;	    	 			     
Col. (3): Satellite used: 'C'=Chandra, 'X'=XMM-Newton,		    				    	 			    
'A'=ASCA; Col. (4): logarithm of the 2-10 keV flux in units of erg cm$^{-2}$ s$^{-1}$, 
not corrected for Compton thick candidates. 
Col. (5): logarithm of the 2-10 keV luminosity, corrected for Compton thick
candidates. Col. (6):
X-RAY REFERENCES: (1) Cappi et al. 2006; (2) This work; (3) Pappa et al. 2001; 
(4) Guainazzi et al. 2005; (5) Terashima et al. 2002; (6) Eracleous et al. (2002); 
Col. (7):  Final classification: '$\surd$'=Compton thick candidates, '$\times$'=Compton thin candidate
and '?'=doubtful objects.; Col. (8): logarithm of the \oiii luminosity corrected for Galactic absorption
and NLR extinction; Col. (9): logarithm of the H${\alpha}$ (broad+narrow component) 
luminosity corrected for Galactic absorption
and NLR extinction; Col. (10): logarithm of the M$_{BH}$ value in units of M$_{\odot}$; Col. (11):
M$_{BH}$ measurement methods; M: maser kinematics;
G: gas kinematics; S: stellar kinematics; R: reverberation mapping; I$_{\sigma}$: 
inferred from the mass-velocity dispersion correlation; Col. (12): M$_{BH}$ REFERENCES:  
(1) Ho (2002); (2) Woo \& Urry (2002); (3) Merloni et al. (2003); (4) Kaspi et al. (2000);
(5) Pellegrini (2005); (6) Merritt \& Ferrarese (2001); (7) di Nella et al. (1995);
(8) Barth et al. (2002); (9) McElroy (1995).}
}			    	 			    
\end{center}					    				    	 			     
\end{table*}				

Other than the $\sim$250 ks of EPIC Guaranteed Time 
for the distance-limited sample of 27 Seyfert
galaxies, we further analyzed 5 public observations
available from the XMM-{\it Newton} Science Archive 
(XSA)\footnote{http://xmm.esac.esa.int/xsa}.
To be consistent with C06, we follow the same
reduction and analysis procedure as described in that paper.
Objects not belonging to the distance-limited survey of C06 are
NGC 1275, NGC 3516, NGC 3227, NGC 5548 and NGC 7479. 

An atlas of \cha and \xmm images and spectra of the sources
has been produced in the 0.3-10 keV energy band (Panessa Ph.D., 2004). 
The results obtained indicate a high detection rate ($\sim$ 95\%)
of active nuclei, characterized, in $\sim$ 60\% of the objects, also by
the presence of nearby off-nuclear sources and/or, 
in $\sim$ 35\% of the objects, diffuse emission.
Altogether these results demonstrate that
high spatial resolution is fundamental for this type of studies in order
to isolate nuclear emission from other X-ray emitting 
components of the host galaxy.

Spectral analysis has been performed in order to first identify the
underlying continuum when possible, then additional components and
features have been included to best reproduce the data
(following the same criteria as in C06). Each spectrum has been 
initially fitted with a model
consisting of a power-law plus absorption fixed at the Galactic 
value and intrinsic absorption left free to vary.
Some high quality spectra required more complex modeling 
and additional features (a soft component and/or emission line features). 
Details on the spectral analysis performed in this 
work are given source by source in the Appendix.
X-ray fluxes have been obtained from the best spectral fit found
and corrected for Galactic and intrinsic absorption (Col. 4, Table~$\ref{x}$), 
except for those cases having 
very poor statistics for which a standard power-law model with a 
photon index fixed at the value of 1.8 and Galactic absorption was assumed.
Luminosities were computed using H$_{\circ}$=75 km s$^{-1}$ Mpc$^{-1}$ 
and distances as in Table~\ref{tab=sample}.

The distributions of spectral parameters, in particular for type 1 objects,
are found to be within the range of values observed in luminous AGN.
The observed distribution
of the instrinsic column densities for the total sample ranges from the typical Galactic
values ($\sim$ 10$^{20}$ cm$^{-2}$) to very high absorptions (i.e. $\sim$ 10$^{23}$ cm$^{-2}$).
Nearly 30\% of type 1 Seyfert galaxies are characterized by a significant 
amount of absorption (i.e. $\geq$ 10$^{22}$ cm$^{-2}$) that could be
ascribed to ionized material and/or dense gas clouds crossing the line of sight. 
The distribution of the column densities found for our type 2 Seyfert galaxies 
deviates from past results showing mostly mildly absorbed 
objects. However, it is well known that Compton thick
sources (with N$_{H}$ $>$ 10$^{24}$ cm$^{-2}$) may appear as objects
with little absorption (Risaliti et al. 1999) if fitted with simple
models and/or if only data with poor statistics are available.
We compared between the column density distributions of type 1.9 and
pure type 2 objects to find no difference (KS probability of 0.42). We
therefore grouped the two classes into a single group.

Summarizing, the X-ray properties of our sample
resemble those of more luminous AGN, except for four
sources which have been only marginally detected in
X-rays (NGC 1058, NGC 3627, NGC 3489 and NGC 4472).
In this work we focus on the nuclear X-ray luminosities 
(Col. 4, Table~$\ref{x}$) which are obtained in a homogeneous way and 
are as less contaminated as 
possible by the presence of diffuse emission
and/or off-nuclear sources. The correction applied to 
the luminosities is 
discussed in the next section.
		     				    	  
\section{How many heavily absorbed sources?}				    	 

\begin{figure}
\begin{center}
\includegraphics[width=0.50\textwidth,height=0.35\textheight,angle=0]{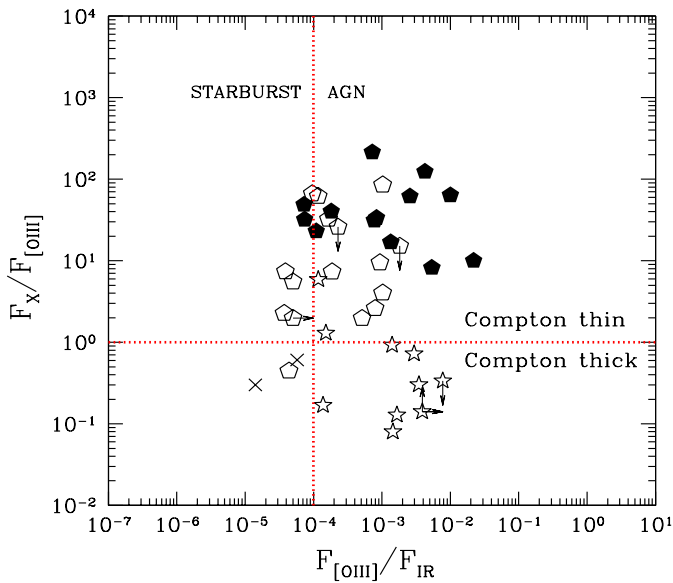}
\caption{F$_{X}$/F$_{[OIII]}$ vs. F$_{[OIII]}$/F$_{IR}$ diagram. 
Compton thin, Compton thick and starburst regions have been separated by dashed lines. 
Type 1 objects are plotted as filled polygons, type 2 as empty polygons, 'mixed Seyfert' objects
as crosses and Compton thick candidates as stars.}
\label{dia}
\end{center}
\end{figure}  

X-ray spectra have proved to be an important tool to 
have a direct estimate of the amount of absorbing 
material in Seyfert galaxies. 
However, for values of the column density $>$ 10$^{24}$ cm$^{-2}$,
X-rays above a few keV are not able to penetrate
the absorbing material and the photoelectric cutoff (if any) in the observed 
spectrum below 10 keV does not provide information on the real column density 
and so the galaxy may be erroneously classified 
as a low-absorption object. 
This leads to an underestimation of the intrinsic hard X-ray luminosity
heavily affecting the shape of the X-ray luminosity
distribution of type 2 Seyferts (C06). 

In order to have luminosities which are as much as possible not affected by absorption,
we have searched within our sample for absorbed objects
not classified as such from the X-ray modelling. To
do so, we have taken advantage of some powerful diagnostic tools
like the flux diagnostic diagrams (Bassani et al. 1999, Panessa \& Bassani 2002). 

Flux diagnostic diagrams are based on 
measuring the X-ray luminosity against an isotropic indicator of the 
intrinsic brightness of the source to evaluate 
the true amount of absorption.
If the UM is correct,  
a molecular torus should be present in Seyfert galaxies that could block the
X-ray emission coming from the central engine when it intercepts
the line of sight. However, 
emission coming from larger scales, like the Narrow Line Region
or a non nuclear starburst region, should not be affected by
obscuration. 
The [OIII]$_{\lambda5007}$ emission line is thought
to originate in narrow line regions by UV ionizing
photons produced by the AGN.
It has been shown in the literature
that the distribution of [OIII]$_{\lambda5007}$ flux
is similar in both types of Seyfert (Mulchaey et al. 1994);
also the NLR size determined from HST data 
is similar in type 1 and type 2 Seyferts (Schmitt et al. 2003).
Although the physics occurring in the NLR is very complex and
the observed luminosity depends on physical and geometrical
properties, for example on the opening angle, 
the [OIII]$_{\lambda5007}$ flux has been extensively used as
an isotropic indicator of the intrinsic AGN power 
(Maiolino \& Rieke 1995; Risaliti et al. 1999;
Bassani et al. 1999, Guainazzi et al. 2005, C06, Heckman 2005).
The relation between the [OIII]$_{\lambda5007}$ and the X-ray
luminosities has also been recently studied
for a sample of broad- and narrow-line Seyfert 1 galaxies (Kraemer et al 2004);
these authors found that the X-ray/[OIII] ratio
can be lowered as a result of the X-ray absorption.
Another frequently used isotropic indicator is the far-infrared emission
(Mulchaey et al. 1994, Mas-Hesse et al. 1994),
probably produced in the coolest regions of the molecular torus
or even over a larger scale; note however that 
infrared emission could be anisotropic
at short wavelengths $<$12 \micron {in heavily absorbed objects}.

The F$_{X}$/F$_{[OIII]}$ and F$_{[OIII]}$/F$_{IR}$
ratio can provide an independent way to establish
which is the dominant component between AGN or starburst and 
at the same time it is a powerful tool in the detection of 
Compton thick sources when an X-ray spectral analysis is not sufficient 
(Panessa \& Bassani 2002).  

A flux diagram is presented in Fig.~\ref{dia}.
All far-infrared fluxes are based on IRAS data derived
from HFS97. The [OIII]$\lambda$5007 flux of each galaxy 
has been corrected for extinction (HFS97). 
Type 1 objects are plotted as filled polygons, 
type 2 as empty polygons and 'mixed Seyferts' as crosses.
Compton thick candidates, as discussed below, have been marked
with stars.
Discriminating values, chosen for SB (starburst), Compton thick
and Compton thin regions, are given as in Panessa \& Bassani (2002).

Most of the sources of both 
types lie in the AGN region and only
a small fraction are located at the boundary between
AGN and Starburst. Indeed, the X-ray analysis
of most of our sample objects has shown
that they host an AGN.
The positions of the few borderline objects 
in our diagram could be due to the low 
IRAS angular resolution\footnote{Note that infrared data
have been taken using a
beam size of $\sim$ 8', while optical and X-ray 
data are derived using much smaller apertures, i.e. $\sim$ 1" and 
$\sim$ 2"-25" respectively.}, 
i.e. the presence of enhanced star-forming
regions in the host galaxy  
may contribute significantly to increase the IR flux in these 
Seyferts; this translates into
F$_{[OIII]}$/F$_{IR}$ ratios
smaller than that of the AGN alone. For example, NGC 5033 is known
to host bright HII regions in the inner parts of its spiral arms 
(P{\' e}rez Garc{\'{\i}}a \& Rodr{\'{\i}}guez Espinosa 1998);
in NGC 4639 (Gonz{\' a}lez-Delgado et al. 1997) 
as well as in NGC 4725 (Sandage \& Bedke 1994) many bright HII regions
have been observed, while NGC 3031 is known to host a very strong 
underlying stellar continuum (Alonso-Herrero et al. 2000).

In order to identify
Compton thick sources, the F$_{X}$/F$_{[OIII]}$ ratio
has been considered: the most populated region is 
that of Compton thin AGN where all broad line Seyferts
and 17 type 2 objects are located while the remaining
type 2 Seyferts lay in the Compton thick region.
As a matter of fact, four out of six known Compton thick sources, 
namely NGC 1068, NGC 1667, NGC 2273 and NGC 5194, occupy 
this region confirming the reliability of this diagnostic diagram,
while NGC 3079, a Compton thick source, is above the Compton thick boundary,
probably because of the presence of filaments in its nuclear region
whose emission contributes significantly in the 0.1-6.5 keV range
(Cecil et al. 2002). 
There is a group of Compton-thick source candidates, namely 
NGC 676, NGC 1167, NGC 3185, NGC 3982 and NGC 7743, for
which X-ray spectral analysis alone cannot draw a conclusion.
No FIR data are available for NGC 3489 (not in the plot), which we classify as
Compton thick from its F$_{X}$/F$_{[OIII]}$ ratio (0.63). 

Objects with an uncertain behaviour constitute a small
fraction of the Seyfert 2 sample (7 out of 34, marked with
a "?" in Table~\ref{x}). Most
of these objects, placed on the SB region of the diagram, 
may even not contain an AGN, actually
NGC 1058, NGC 3627 and NGC 3486 only have
upper limits in X-rays. 

In summary, a sub-sample of 11 secure Compton thick candidates have 
been recognized (they have been marked with a $\surd$ in Col. 6
of Table~\ref{x}); these objects are all Compton thick
according to the flux diagnostic diagram and in five cases
this indication is supported by a strong Fe line of a $>$ 1 keV
equivalent width, i.e. NGC 1068 (Matt et al. 1997), NGC 1667 (Bassani et al. 1999), NGC 2273
(Maiolino et al. 1998), NGC 3079 (Cecil et al. 2002) and 
NGC 5194 (Terashima \& Wilson 2001)\footnote{Note that in the other six cases,
the non detection of an Fe line could be due to
the poor statistics of the X-ray spectra.}.

These tools have revealed
that the fraction of objects which may be affected by Compton thick  
obscuration among type 2 Seyferts ranges from 30\% (if only the secure Compton thick sources are considered) 
up to 50\% (if also objects with an uncertain behaviour are 
Compton thick sources) in agreement with previous 
estimates available for a flux-limited sample (Risaliti et al. 1999)
and a distance limited sample (C06).

\begin{figure*}
\begin{center}
\parbox{18cm}{
\includegraphics[width=0.50\textwidth,height=0.35\textheight,angle=0]{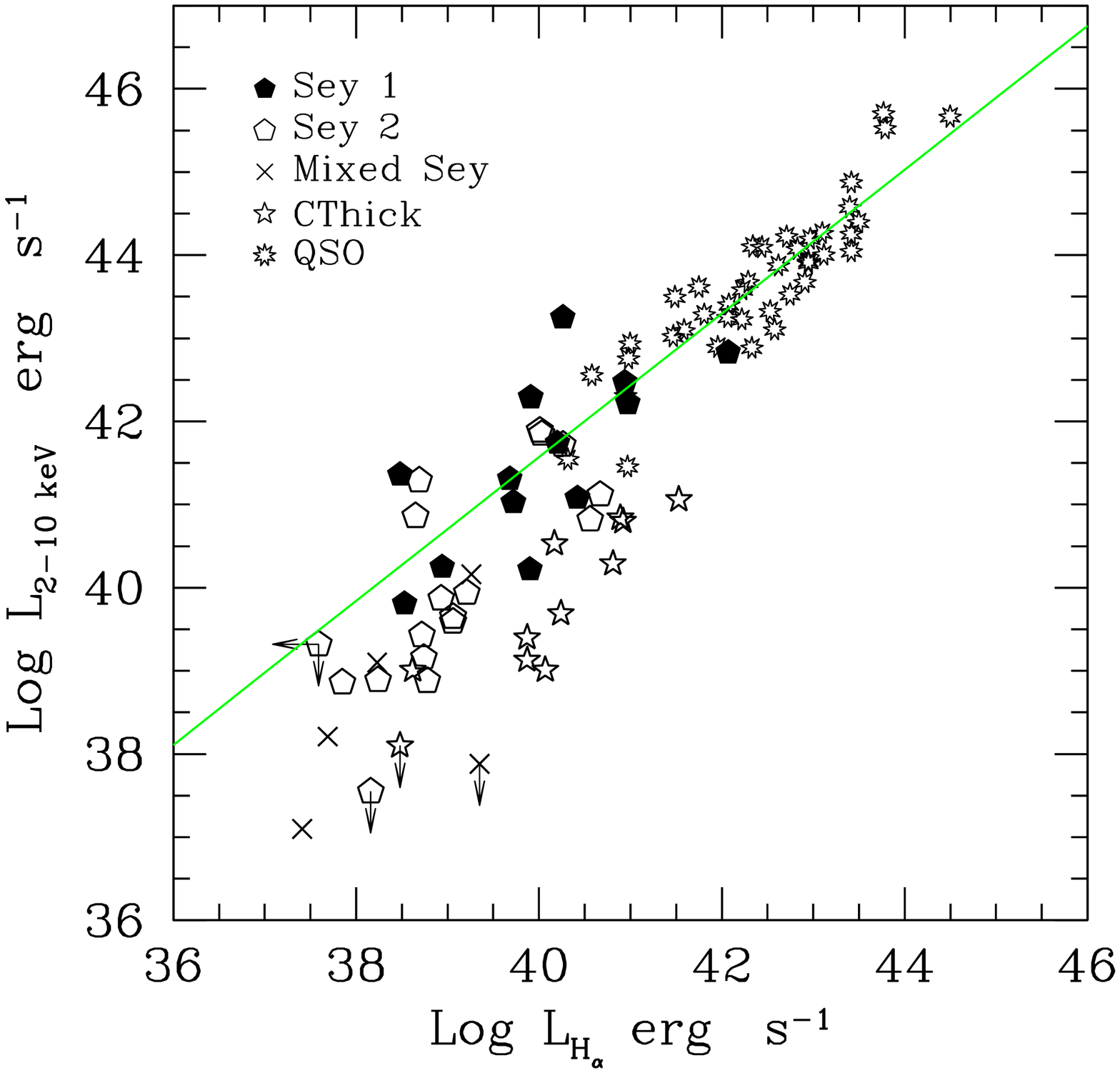}
\includegraphics[width=0.50\textwidth,height=0.35\textheight,angle=0]{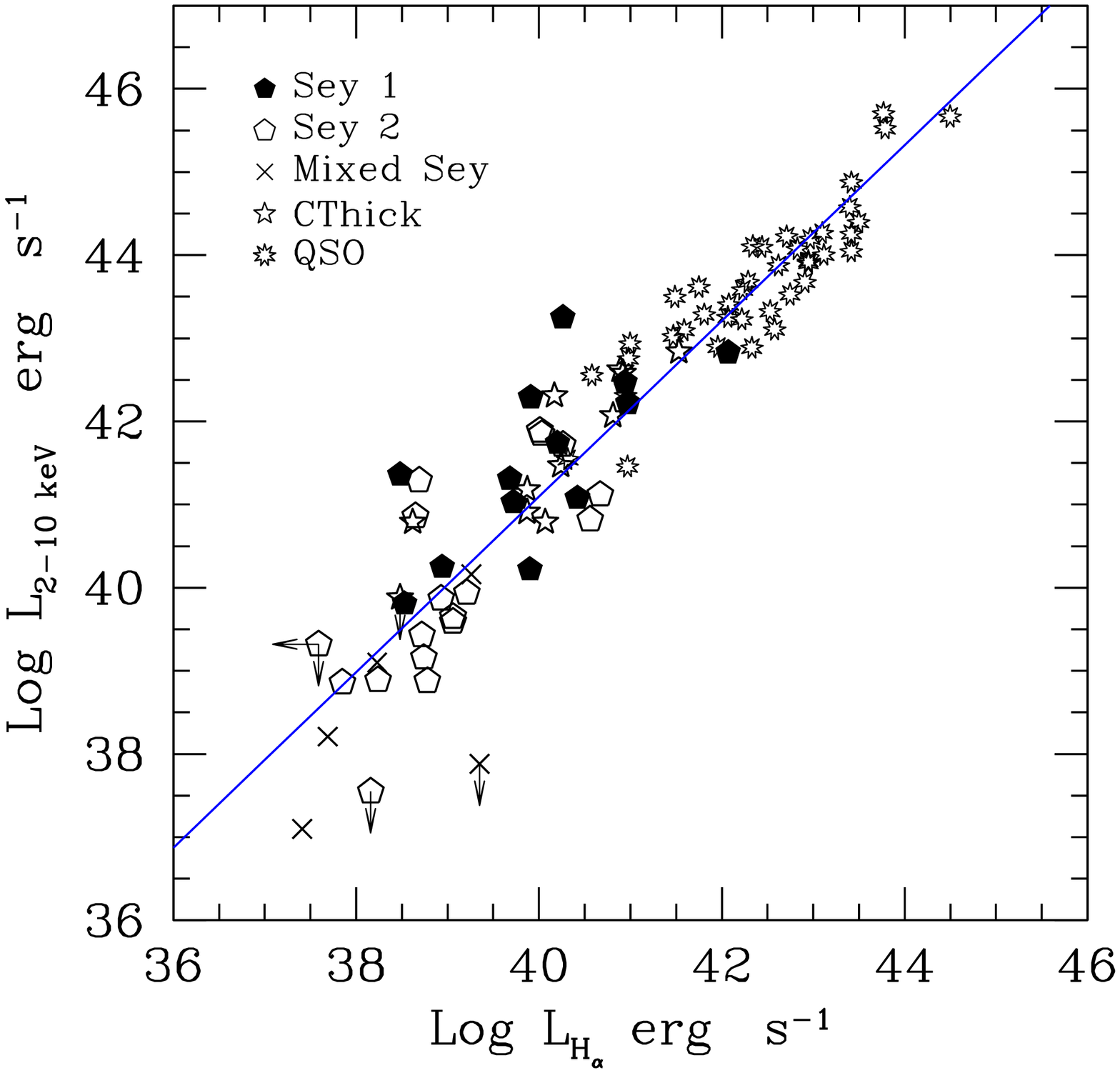}}
\caption{Left panel: Log of 2-10 keV luminosity
versus log of H$_{\alpha}$ luminosity (narrow+broad components) corrected
for the Galactic and NLR extinction. The dotted line
shows the best fit linear regression obtained by fitting type 1 Seyferts
and low-z quasars from Ward et al. (1988). Right panel: 
the same plot using the 'Compton thick' corrected
luminosity. The solid line shows our best fit linear regression line
obtained by fitting the total sample of Seyfert galaxies and the
low-z quasars.
Filled polygons are type 1 Seyfert, open polygon sare type 2 Seyfert,
'mixed Seyfert' objects are indicated as crosses, Compton thick candidates
are stars and low-z quasars are 'rounded-stars'.}
\label{al}
\end{center}
\end{figure*}  

The X-ray luminosities of Compton thin sources (both type 1 and type 2
Seyferts) have been corrected for the column density measured from the
2-10 keV spectra.
Assuming that all the secure candidates are actually Compton thick, then their
intrinsic column density should be higher than $\sim$ 10$^{24}$ cm$^{-2}$.
This prevents any possibility to directly measure 
the intrinsic X-ray luminosity in the 2-10 keV band,
since the latter is completely blocked by the absorbing matter
and we can only measure a reflected/scattered component, if any.
Therefore, to estimate the intrinsic X-ray luminosity
we corrected the observed one by a factor of 60.
This value has been obtained by means of 
the F\hard/F$_{[OIII]}$ ratio: the mean value of this ratio has been
calculated for our type 1 objects (Log~$<$F\hard/F$_{[OIII]}$$>$~$\sim$~1.74 $\pm$ 0.49)
and for the five known Compton thick sources 
(Log~$<$F\hard/F$_{[OIII]}$$>$~$\sim$ -0.03 $\pm$ 0.07). 
The difference between these mean values can be considered as an
approximate value of the correction factor when estimating the intrinsic
luminosity of our Compton thick sources, i.e, $<$F\hard/F$_{[OIII]}$$>$$_{Type
1}$/$<$F\hard/F$_{[OIII]}$$>$$_{Type 2, CThick}$~$\sim$~60,
hereafter we refer to 'Compton thick' corrected
X-ray luminosities (see also C06).
In Table~\ref{x}, X-ray luminosities for the sample sources 
have been corrected for the intrinsic absorption measured from their X-ray spectra, 
while for Compton thick candidates we applied the above correction factor.
Although the latter should be considered as an indicative value,
it is however in tune with what observed with BeppoSAX
in Circinus Galaxy (Matt et al. 1999) and
NGC 4945 (Guainazzi et al. 2000).
Moreover, in the case of NGC 1068, 
Iwasawa, Fabian, \& Matt (1997) estimate 
the intrinsic luminosity to be in the range
of 10$^{43}$ - 10$^{44}$ erg/s which is a factor of 100-1000 higher than
the observed luminosity. This large correction factor is also
theoretically justified by the torus model proposed by 
Ghisellini, Haardt \& Matt (1994),
in which they derive, a correction factor
ranging from 100 to 2$\times 10^{3}$ depending on the amount of
absorption and on the viewing angle with respect to the
obscuring torus.
Finally, it is worth noticing that before correcting
the X-ray luminosity for the Compton thick sources,
type 1 and type 2 Seyfert galaxies show a different
distribution in luminosities (with KS test of 0.001),
type 2 objects have lower luminosities than
type 1 objects. After correcting for Compton thick sources, 
the KS probability is reduced to 0.05, confirming
the previous finding by C06.
	     
\section{X-ray vs Optical Emission Line Luminosity Correlations}
 
The detection of an X-ray nucleus in almost all
our sources is a strong evidence in favour
of the presence of an AGN even at very low
luminosities.
Another convincing way to investigate 
the activity of the central
source is through the observed correlations
between X-ray and optical emission line luminosities since
a proportionality between 
these quantities is expected in AGN. In 
luminous sources strong correlations between
\ha, H$_{\beta}$, [OIII]$_{\lambda5007}$ luminosities and X-ray luminosities
have already been found (Elvis, Soltan \& Keel 1984, 
Ward et al. 1988, Mulchaey et al. 1994). 

It is worth investigating whether
this correlation holds in our sample, in order
to further test whether  
the same physical processes occurring in luminous AGN are also
in action in their low luminosity counterparts.

To characterize quantitatively an apparent correlation
between two properties of the sample under study we apply a
linear regression statistical procedure. We have used the EM 
(Expectation-Maximization) algorithm 
since it deals with censored data; this algorithm is implemented
in the ASURV package (Isobe, Feigelson \& Nelson 1986).

Table~\ref{stat} shows 
the statistical properties of the correlations.
The sample has been grouped in sub-classes (type 1 and type 2);
quasar samples (as described below) have been added to the
analysis.

\subsection{X-ray vs. H$_{\alpha}$ luminosity}

A strong positive correlation between the X-ray and
the H$_{\alpha}$ emission line luminosity 
has been widely observed in classical AGN,
such as Seyferts and quasars (Ward et al. 1988)
and in low luminosity AGN, such as LINERs (Terashima,
Ho \& Ptak 2000, Ho et al. 2001). It has been
shown that this correlation is not
an artifact of distance effects. Typical ratios
observed in bright objects are Log (L$_{X}$/L$_{H\alpha}$) $\sim$ 1-2,
supporting the idea that optical 
emission lines arise in gas photoionized by
the central nucleus.

In Fig.~\ref{al} the
logarithm of the 2-10 keV luminosity has been plotted versus
the log L$_{H\alpha}$, the latter including both
the narrow and broad (if present) components of the line,
corrected for extinction due to the Galaxy and to the narrow-line region
(Ho, Filippenko \& Sargent 1997).
The X-ray luminosities are from this
work (Table~\ref{x}) without (left panel) and with (right panel)
correction applied to Compton thick candidates.
A sample of low-z quasars from  Ward et al. (1988)
has also been included to compare our results with
high luminosity objects (luminosities have been 
adjusted to H$_{0}$= 75 km s$^{-1}$ Mpc$^{-1}$).
Each class of AGN has been marked with different symbols:
filled polygons are type 1 Seyfert, open polygons are type 2 Seyferts,
crosses are 'mixed Seyferts', stars are the Compton thick candidates
and rounded-stars are low-z quasar objects.

\begin{figure}
\begin{center}
\includegraphics[width=0.50\textwidth,height=0.35\textheight,angle=0]{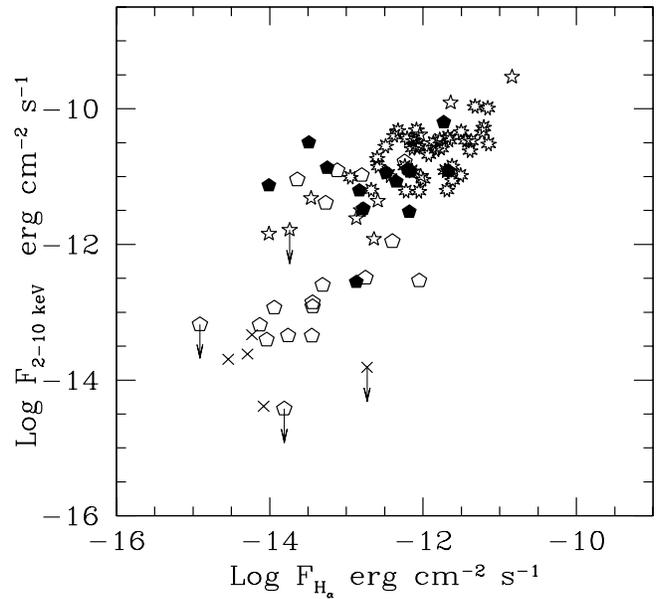}
\caption{Correlation between 2-10 keV versus H$_{\alpha}$ fluxes. 
H$_{\alpha}$ narrow+broad components are corrected
for the Galactic and NLR extinction. Symbols as in Fig.~\ref{al}}
\label{axflux}
\end{center}
\end{figure}  

\begin{figure*} 
\begin{center}
\parbox{18cm}{
\includegraphics[width=0.50\textwidth,height=0.35\textheight,angle=0]{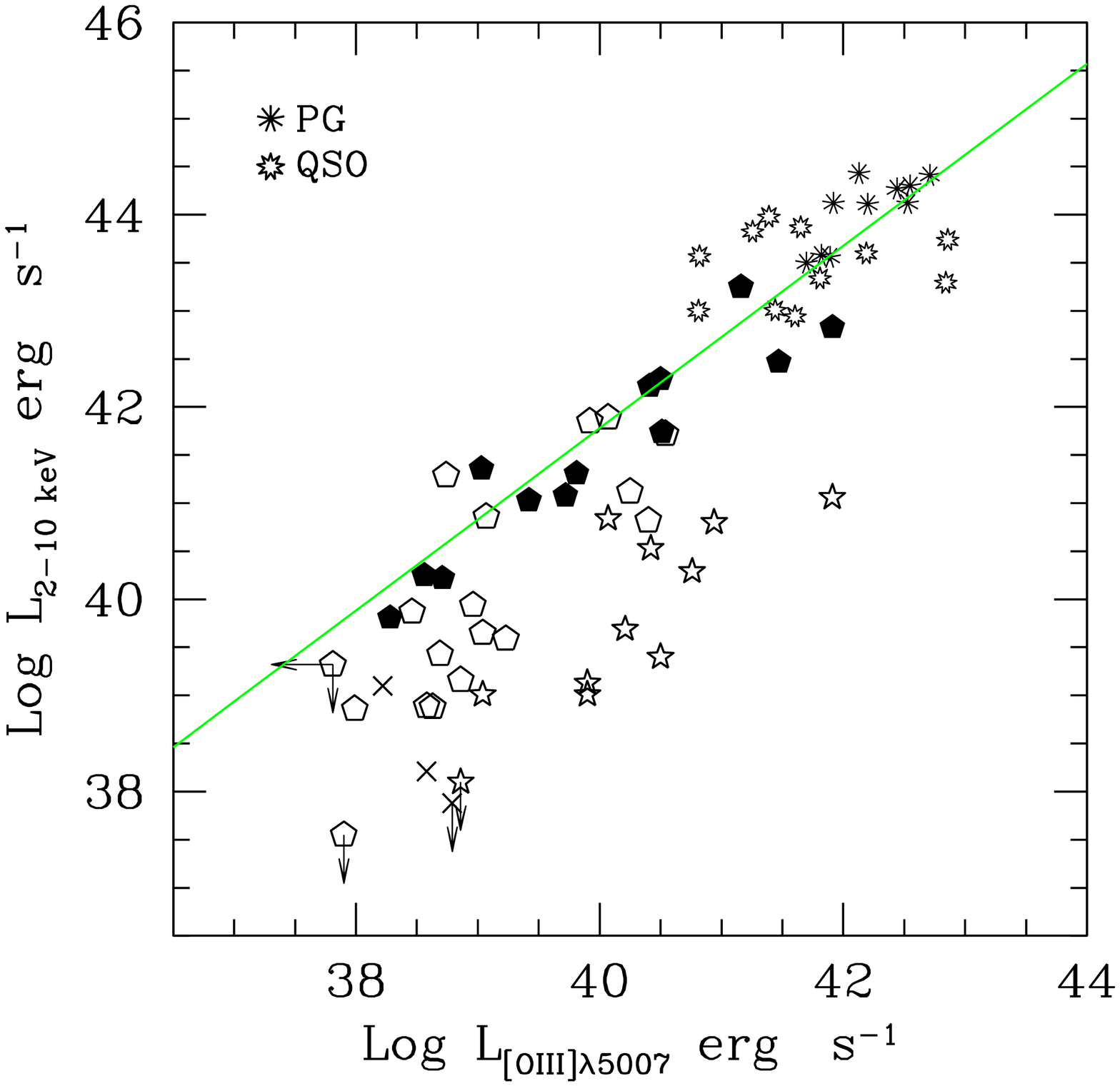}
\includegraphics[width=0.50\textwidth,height=0.35\textheight,angle=0]{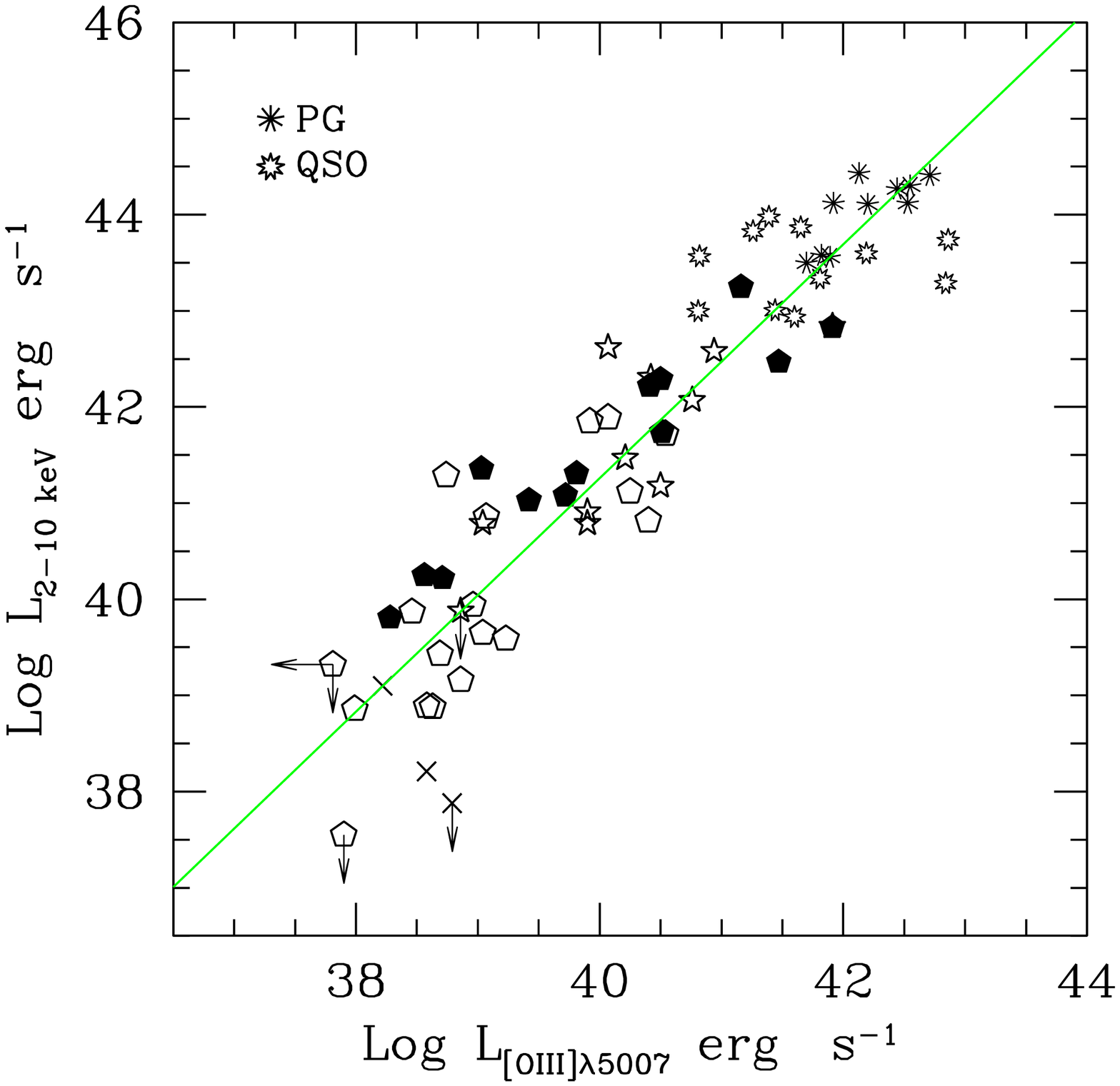}}
\caption{Left panel: Log of 2-10 keV luminosity
versus log of [OIII]$_{\lambda5007}$ luminosity corrected
for the Galactic and NLR extinction. The solid line
shows the best fit linear regression line obtained by fitting
type 1 Seyfert,  a sample of bright type 1 Seyferts (QSO, Mulchaey et al. 1994) and
a sample of PG quasar (PG, Alonso-Herrero et al. 1997).
Right panel: the same plot using 'Compton thick' corrected
luminosity. The solid line shows our best fit linear regression line
for the total sample of Seyfert galaxies, the PG quasars
and bright type 1 Seyfert. Symbols are as in Fig.~\ref{al}}
\label{ox}
\end{center}
\end{figure*}  

First, we checked whether a correlation is present
among our Seyfert sample. As reported in Table~\ref{stat},
a strong correlation is found in the sample.
We fitted type 1 Seyferts and low-z quasar 
objects in order to compare objects
with broad+narrow component of the H$_{\alpha}$ line.
The solid line in the left panel represents our 
best-fit linear regression line. This correlation
is highly significant (Spearman's rho=0.93),
although type 1 Seyferts in our sample
show a larger scatter than the low-z quasars.
Note that the X-ray and H$_{\alpha}$ data have not been taken
simultaneously. Consequently, the strong X-ray flux variability observed
in most type 1 Seyferts (NGC3516, Guainazzi et al. 2001;
NGC 4051, Lamer et al. 2003; NGC4151, Czerny et al. 2003, 
NGC4395, Iwasawa et al. 2000;) is a likely source
of scatter. At lower luminosities, 
type 2 objects and 'mixed Seyferts' show 
L$_{X}$/L$_{H\alpha}$ ratios lower than type 1s and low-z quasars.
Note, however, that we are comparing type 1 objects
which have both narrow and broad H$_{\alpha}$ line components 
and type 2 Seyferts which lack the broad H$_{\alpha}$ component. 

On the right panel of Fig.~\ref{al}, the 2-10 keV luminosity
of Compton thick candidates has been corrected
and the solid line is our best-fit regression line obtained by fitting
the total Seyfert sample and the low-z quasar sample. 
This regression line is consistent with that
found by Ho et al. (2001) using the same low-z quasar sample
and a sample of LLAGN observed with {\it Chandra}. 

The correction in the X-ray luminosity significantly reduces
the scatter at low luminosities and the inclusion of 
type 2 Seyferts in the calculation of the
regression line results in a steeper slope with respect to 
those of type 1 and low-z quasars. 
The steepening of the regression line is probably enhanced by
those Compton thin type 2 objects at low luminosities,
which show an optical excess emission with respect to the
X-ray luminosity.
The latter could be due to the contribution given by a circumnuclear
starburst to the H${\alpha}$ emission which is more
important in low luminosity sources than in high
luminosity ones, where the emission is completely dominated by the AGN.

The same correlation using fluxes
is shown in Fig.~\ref{axflux} (rho=0.78, prob $<$ 0.001) 
strongly supporting that the 
luminosity-luminosity correlation is not due to distance effects.

It is remarkable that the 
present sample traces the L$_{X}$ - L$_{H\alpha}$ relation for 
luminosities typical of luminous AGN (10$^{41}$ - 10$^{43}$ erg s$^{-1}$)
down to low luminosities typical of LINERs (10$^{37}$ - 10$^{40}$ erg s$^{-1}$).
The strong correlation observed between L$_{X}$ - L$_{H\alpha}$
suggests that the dominant ionization source in our sample
is photoionization by an AGN and that optical/UV photons
are somehow linked to the X-ray emission.
A few objects at 2-10 keV luminosities of 
$\sim$ 10$^{38}$ erg sec$^{-1}$
appear to be systematically under the correlation.
Three out of four of these sources are actually those 
classified as 'mixed Seyfert', i.e. NGC~2683, NGC~3627 and NGC~6503;
moreover for NGC 1058 and NGC 3627 only an upper limit to the X-ray luminosity is
available.
If the ionization mechanism of the emission lines
is due to or enhanced by hot stars or shocks, 
the resulting L$_{X}$/L$_{H\alpha}$ ratio is expected to be accordingly
smaller compared to 
those of AGN. Indeed, P{\' e}rez-Olea \& Colina (1996) 
have shown how the 
L$_{X}$/L$_{H\alpha}$ ratio in AGN is larger than
in starbursts up to a factor of $\sim$100 independently
on the intrinsic luminosity or activity level.
This could be a possible scenario for these sources which,
from our diagnostic diagrams, have been recognized 
as our most probable starburst candidates. However
we cannot rule out that a very faint/heavily obscured AGN is present in such
objects, as for example in the 'Elusive AGN' in which the nucleus is heavily obscured
and there are no optical evidences for the presence
of an AGN (Maiolino et al. 2003).

\begin{figure}[!htb] 
\begin{center}
\includegraphics[width=0.50\textwidth,height=0.35\textheight,angle=0]{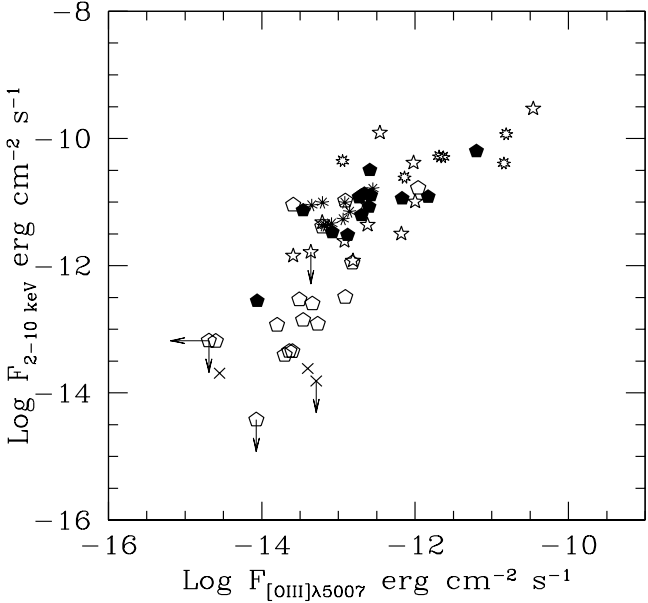}
\caption{Correlation between log of 2-10 keV versus log of \oiii fluxes, the latter
corrected for the Galactic and NLR extinction.}
\label{oxflux}
\end{center}
\end{figure}  

\begin{table*}
\small{
\caption{\bf Correlation statistics in luminosities}
\label{stat}
\begin{center}
\begin{tabular}{lcccccccc}
\hline
\hline
\multicolumn{1}{c}{Variables} &
\multicolumn{1}{c}{Sample} &
\multicolumn{1}{c}{N} &
\multicolumn{1}{c}{X(ul)} &
\multicolumn{1}{c}{Y(ul)} &
\multicolumn{1}{c}{rho} &
\multicolumn{1}{c}{Prob.} &
\multicolumn{1}{c}{a} &
\multicolumn{1}{c}{b} \\
\multicolumn{1}{c}{(1)} &
\multicolumn{1}{c}{(2)} &
\multicolumn{1}{c}{(3)} &
\multicolumn{1}{c}{(4)} &
\multicolumn{1}{c}{(5)} &
\multicolumn{1}{c}{(6)} &
\multicolumn{1}{c}{(7)} &
\multicolumn{1}{c}{(8)} &
\multicolumn{1}{c}{(9)}\\
\hline
\hline
Log L$_{X}$ vs Log L$_{H\alpha}$  &  Tot  &	 47  & 1  &   4 & 0.82  &$<$0.001   & 1.20$\pm$0.12 &   -6.82$\pm$4.82 \\
Log L$_{X}$ vs Log L$_{H\alpha}$  &  S1   &    13  & 0  &	0 & 0.68  &   0.019 & 0.74$\pm$0.21 &   12.12$\pm$8.34 \\
Log L$_{X}$ vs Log L$_{H\alpha}$  &  S2   &    34  & 1  &	4 & 0.83  &$<$0.001 & 1.28$\pm$0.14 &  -10.04$\pm$5.61 \\
\\
Log L$_{X}$ vs Log L$_{H\alpha}$  &  Tot+QSO$^{a}$ & 87  & 1  &   4 & 0.95  &$<$0.001 & 1.06$\pm$0.04 & -1.14$\pm$1.78 \\
Log L$_{X}$ vs Log L$_{H\alpha}$  &  S1+QSO$^{a}$  & 53  & 0  &   0 & 0.93  &$<$0.001 & 0.86$\pm$0.05 &  6.98$\pm$2.08\\
\\
\\
Log L$_{X}$ vs Log L$_{[OIII]}$	   &  Tot   & 45  & 1  &   4 & 0.88 &$<$0.001 & 1.22$\pm$0.11 &   -7.55$\pm$4.33 \\
Log L$_{X}$ vs Log L$_{[OIII]}$	   &  S1    & 13  & 0  &   0 & 0.93 &$<$0.001 & 0.86$\pm$0.09 &    6.99$\pm$3.51 \\
Log L$_{X}$ vs Log L$_{[OIII]}$	   &  S2    & 32  & 1  &   4 & 0.86 &$<$0.001 &  1.34$\pm$0.15 &  -12.32$\pm$5.80 \\
\\
Log L$_{X}$ vs Log L$_{[OIII]}$	 &  Tot+QSO$^{b}$ & 66  & 1  &   4 & 0.93 &$<$0.001 & 1.22$\pm$0.06 &  -7.34$\pm$2.53   \\
Log L$_{X}$ vs Log L$_{[OIII]}$	 &  S1+QSO$^{b}$  & 34  & 0  &   0 & 0.83 &$<$0.001 & 0.95$\pm$0.07 &  3.87$\pm$2.76   \\
\hline
\end{tabular}
\end{center}
Notes: Statistical properties of the 'Compton thick' corrected 2-10 keV X-ray luminosity 
versus H$_{\alpha}$ and [OIII]$\lambda$5007 luminosities; Col. (1): Variables X and Y; 
Col. (2): Subsample considered: Tot = total Seyfert sample, 
S1 = Seyfert 1 galaxies, S2 = Seyfert 2 galaxies; QSO$^{a}$= low-z quasars (Ward et al. 1988);
QSO$^{b}$= bright type 1 Seyfert (Mulchaey et al 1994) + PG quasars (Alonso-Herrero et al. 1997);
Col. (3): number of objects; Col. (4)-(5): Number of upper limits in variable X and Y;
Col. (6-7) Spearman's rho correlation coefficient and the
associated probability P for accepting the null hypothesis
that there is no correlation; Col. (8)-(9):
Correlation coefficient of the best fit linear regression line calculated using EM algorithm
(Isobe, Feigelson \& Nelson 1986), Y=a$\times$ X + b. 
}
\end{table*}
										
\subsection{X-ray vs. [OIII]$_{\lambda5007}$ luminosity}

As previously discussed in Sec.~4, 
the [OIII]$_{\lambda5007}$ emission line is 
thought to be an isotropic indicator and should,
therefore, be representative of the intrinsic power of the
central engine. Although the properties 
of [OIII]$_{\lambda5007}$ have been widely
investigated in the literature, the L$_{X}$ vs. L$_{[OIII]}$ 
correlation itself has been poorly studied. 
Mulchaey et al. (1994)
have found a strong correlation between 
the emission line flux and the observed UV continuum
in type 1 Seyferts, while no correlation was observed
in type 2 Seyferts. This finding is in agreement with the torus model,
where the UV photons are scattered while the line emission
is produced by a directly viewed component.
However these authors made use
of an heterogeneous sample
and did not take into account the presence
of Compton thick sources.
More recently, Heckman et al. (2005) have found that
L$_{X}$ and L$_{[OIII]}$ are well correlated in a sample
of hard X-ray selected AGN (both type 1 and type 2). 
They also found that for a sample of [OIII] flux selected sources, 
the correlation L$_{X}$ vs. L$_{[OIII]}$ for type 1 objects
is consistent with that found in the type 1 hard X-ray sample.
However, the type 2 objects spread a much wider range in the luminosity ratio
and many of them are very weak in hard X-rays.

In Fig.~\ref{ox} we show the 2-10 keV luminosity (without
correction for Compton thick sources, left panel,
and with correction, right panel)
versus the [OIII]$_{\lambda5007}$ luminosity (corrected
for the Galactic and NLR extinction, Ho, Filippenko \& Sargent 1997).
Two comparison samples of bright AGN have been included
in the analysis chosen for having both X-rays and [OIII]$_{\lambda5007}$ 
fluxes available: 
1) a sample of luminous type 1 Seyfert galaxies (hereafter QSO)
from Mulchaey et al. (1994); 2)
a sample of PG quasars (hereafter PG) from Alonso-Herrero et al. (1997).
Luminosities have been adjusted to H$_{0}$= 75 km s$^{-1}$ Mpc$^{-1}$.
The two chosen samples of luminous AGN are not meant to be
complete and biases against low luminosity objects
are probably introduced. However the low luminosity ranges are covered
by our sample and they are just taken as representatives of
the class of luminous sources.

The solid line in the left panel of Fig.~\ref{ox} 
is the best fit linear regression 
line obtained by fitting type 1 Seyferts, 
the QSO sample and the PG sample. 
Contrary to what is obtained in the L$_{X}$ vs. L$_{H\alpha}$
relation, most of the scatter here is introduced by QSOs,
while type 1 Seyferts of our sample follow a tighter correlation.
There is a clear separation between type 1 and type 2 Seyferts.
Once we correct the X-ray luminosity assuming the presence of
'Compton thick' sources, the correlation between the
two luminosities is tighter as shown in the 
right panel of Fig.~\ref{ox}
in which the best fit linear regression line
for the total sample, the QSO and PG samples is over-plotted.
As also found in the L$_{X}$ vs. L$_{H\alpha}$ correlation, 
the slope of the regression line is steeper with respect to 
those obtained for only type 1 and low-z quasars. At low luminosities, 
a possible contribution by a circumnuclear starburst
to the [OIII] emission could explain the excess of the [OIII]
luminosity with respect to the X-ray one observed in a few sources.
Actually, at very low luminosities, the same sources which lay below the correlation
in the X-ray vs H$_{\alpha}$ plot,
lay also below the L$_{X}$ vs. L$_{[OIII]}$ relation
(e.g., NGC~1058, NGC~2683 and NGC~3627). 
The X-ray versus [OIII] correlation still holds in the flux-flux 
plot of Fig.~\ref{oxflux} (rho=0.78, Prob $<$ 0.001). 
As for the H$_{\alpha}$ luminosity, also the [OIII]$_{\lambda5007}$ flux
appears to be a good tracer of the AGN power,
and both correlations are good tools to estimate the 
expected X-ray luminosity. The following relations are obtained
for our sample:

\begin{center}
log L$_{X}$ = (1.06 $\pm$ 0.04) log L$_{H\alpha}$ + (-1.14 $\pm$ 1.78)
\end{center}

\begin{center}
log L$_{X}$ = (1.22 $\pm$ 0.06) log L$_{[OIII]}$ + (-7.34 $\pm$ 2.53)
\end{center}
 
\section{M$_{BH}$ and Eddington ratios}

\begin{figure*} 
\begin{center}
\parbox{18cm}{
\includegraphics[width=0.50\textwidth,height=0.35\textheight,angle=0]{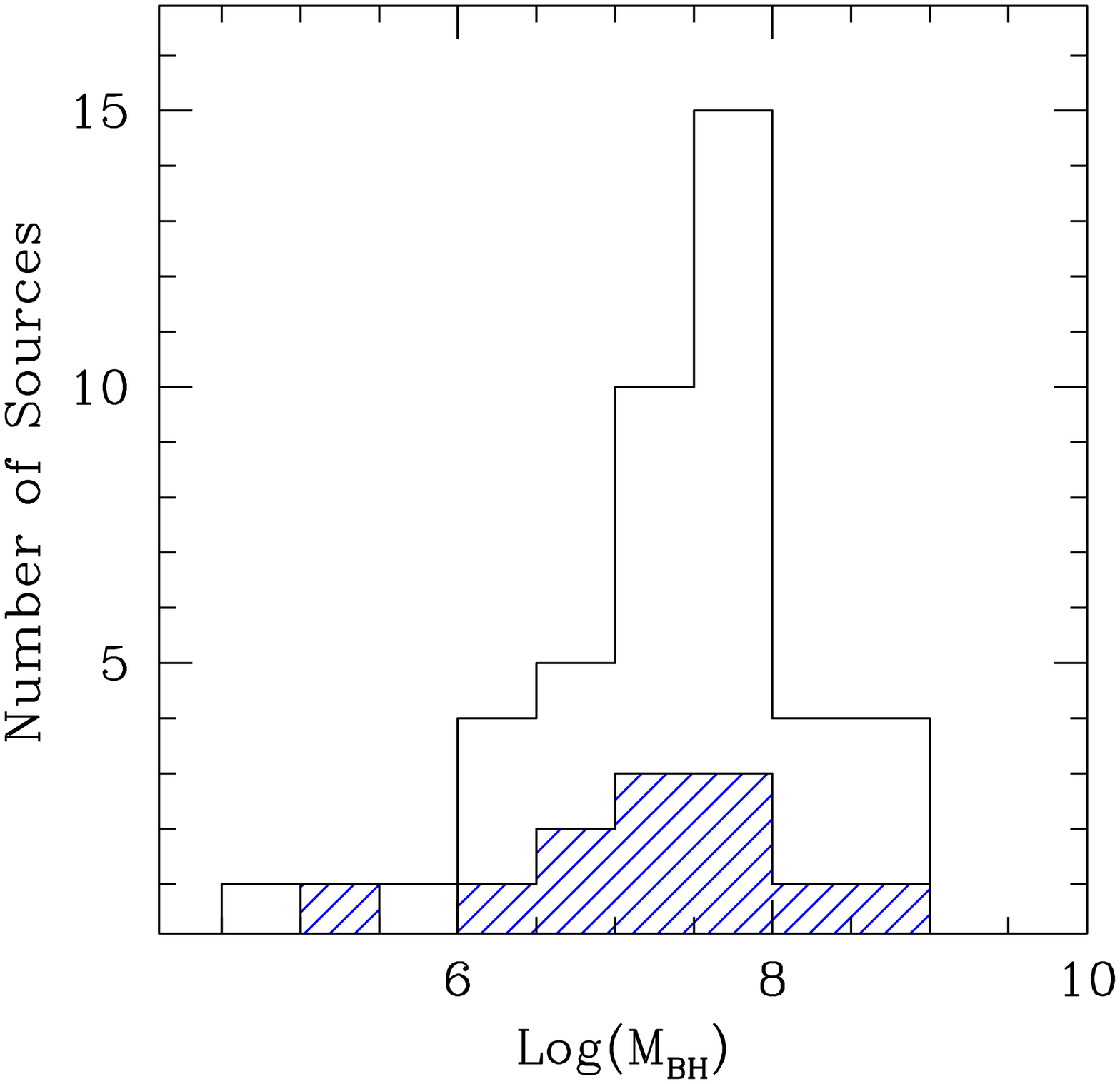}
\includegraphics[width=0.50\textwidth,height=0.35\textheight,angle=0]{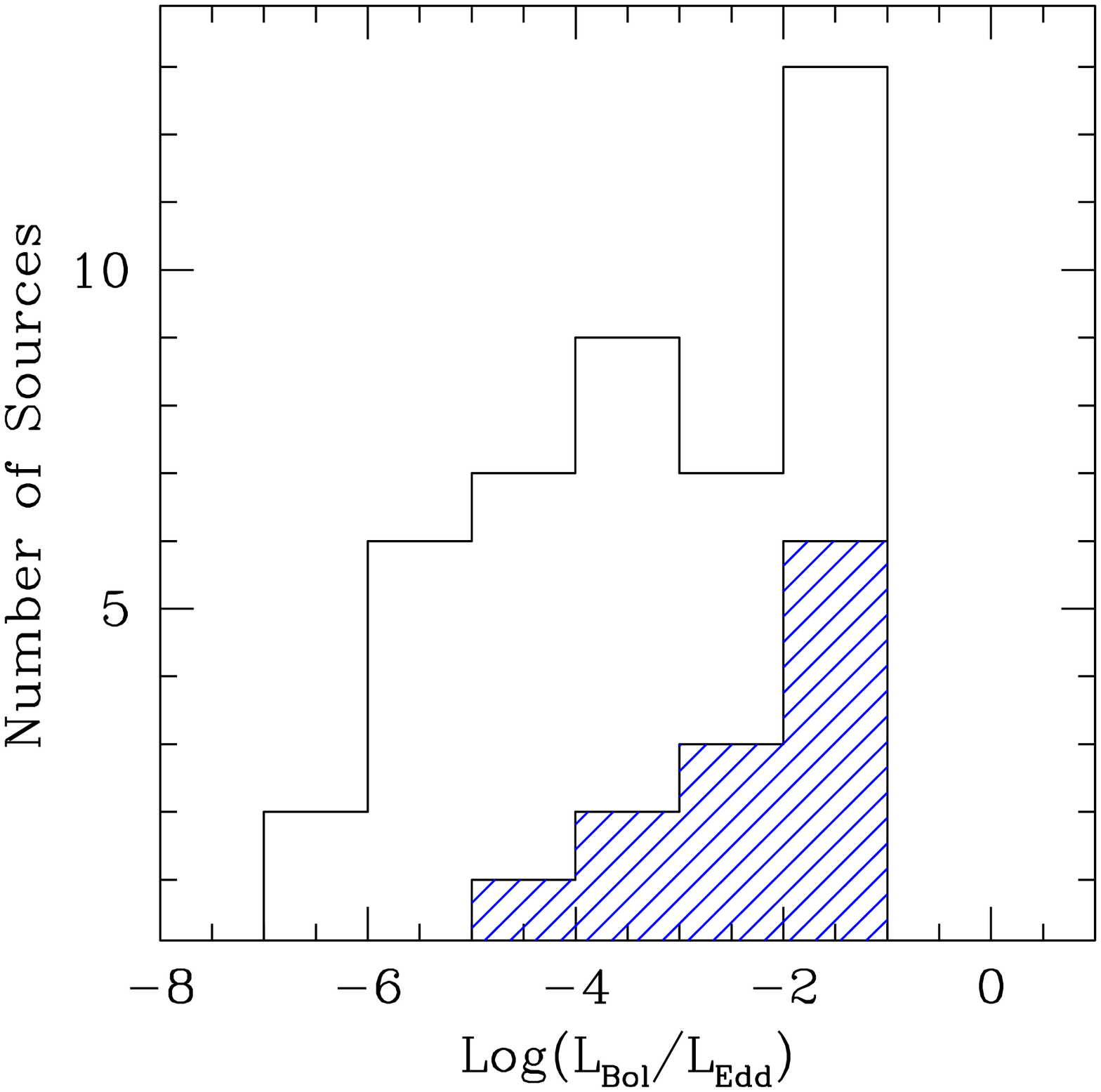}}
\caption{Left panel: Distribution of the log of black hole masses
in unit of solar masses . 
Right panel: Distribution of the log of L$_{Bol}$/L$_{Edd}$ ratio,
assuming that L$_{Bol}$/L$_{X}$ $\sim$ 30.
In both panels the shaded areas represent the distribution of 
type 1 Seyferts only.}
\label{mbh}
\end{center}
\end{figure*}

Black hole mass (M$_{BH}$) estimates are available
for 44 out of 47 objects in our sample (col 9,
Table~\ref{x}). The M$_{BH}$ 
from the literature, have been estimated
in different ways from gas, stellar and maser kinematics
to reverberation mapping or inferred from
the mass-velocity dispersion correlations
(Ferrarese 2002, Tremaine et al. 2002).
For a group of objects stellar
velocity dispersions are available from di Nella et al. (1995),
Barth et al. (2002) and McElroy (1995),
and we calculate the M$_{BH}$ using the
Tremaine et al. (2002) relation.
In Table~\ref{x} we report all these M$_{BH}$
estimates, the method used to calculate them and 
the corresponding reference.

As shown in Fig.~\ref{mbh} (left panel),
black hole masses are fairly sampled 
from $\sim$ 10$^{5}$ to 10$^{8}$ M$_{\odot}$
with a peak at 10$^{7-8}$ M$_{\odot}$;
this figure further indicates that type 1 and type 2 {Seyferts} 
show a similar distribution.
Woo \& Urry (2002) presented a large
compilation of black hole masses, for
an heterogeneous sample
of 234 AGN, ranging from $\sim$ 10$^{6}$ to 
$\sim$ 10$^{10}$ M$_{\odot}$. 
From a comparison of different methods
in estimating M$_{BH}$, they found 
that M$_{BH}$ values estimated from 
reverberation mapping and stellar velocity
dispersion are those more reliable.
Uncertainties in M$_{BH}$ estimates 
for our sample are probably introduced
by the scatter in the  
mass-velocity dispersion correlations and are typically
of the order of 0.3-0.5 dex.
We refer to the above mentioned papers
for a detailed description of the different
methods used to estimate M$_{BH}$
and the relative errors associated
to these measurements. 

The L$_{Bol}$/L$_{Edd}$ ratio
distribution is plotted in the right panel of Fig.~\ref{mbh}.
To calculate the bolometric luminosity we have 
assumed that L$_{Bol}$/L$_{2-10 keV}$ $\sim$ 30.
The latter value is typical of luminous AGN,
being normally in the range of 25-30 (Risaliti \& Elvis 2004, Elvis et al. 1994).
However, it must be kept in mind that the bolometric luminosity here is just
a multiple of the X-ray luminosity, while it really depends on the shape
of the spectral energy distribution which could differ among 
high and low luminosity AGN (Ho 1999, Marconi et al. 2004); actually, 
in LLAGN, the observed L$_{Bol}$/L$_{2-10 keV}$ ratio ranges from 3 to 16 (Ho 1999).
The L$_{Bol}$/L$_{Edd}$ ratio distribution for the total Seyfert sample
covers a wide range of Eddington ratios going down to 10$^{-7}$.
If a L$_{Bol}$/L$_{2-10 keV}$ ratio of 10 is considered,
then Eddington ratios would be a factor of 3 lower than those in Fig.~\ref{mbh}.
The distribution of type 1 objects has been marked
with a shaded region. It is interesting to note that 
a Kolmogorov-Smirnov test gives a probability of 0.01 
for the type 1 and type 2 sub-samples of being
drawn from the same parent population, suggesting that
type 2 objects are accreting at lower Eddington ratios
with respect to type 1 Seyfert galaxies. However, there are some caveats
to take into account: i. the KS probability
of the type 1 and type 2 X-ray luminosity distribution is 0.05,
consequently affecting the bolometric luminosity distributions;
ii. as already pointed out, the bolometric correction could be not a constant,
depending for example on luminosity and therefore could
change from object to object;
iii. objects with an uncertain behaviour, those
in which the star-formation probably dominates,
have been included in the type 2 class. As a matter of fact,
if 'mixed Seyferts' are excluded from the
type 2 sub-sample, the KS probability of the
L$_{Bol}$/L$_{Edd}$ ratio distributions is 0.05.
Moreover, it is possible that with the present data
we are not able to detect all the absorbed low luminosity type 2s,
e.g. Compton thick sources not recognized to be such 
by the diagnostic tools and/or
objects with part of the X-ray emission absorbed by
parsec scale clumpy material detected in a low N$_{H}$ state (e.g., NGC 4388, Elvis et al. 2004).  
However, if a trend of type 2 Seyfert galaxies having 
lower Eddington ratios than type 1 ones is present,
this would have several interesting implications:
for example it would nicely confirm a model 
proposed by Nicastro (2000) that relates the formation
of the broad emission lines of active galactic nuclei
to the accretion rates, i.e. for very low accretion rates 
the BLR would no longer exist. More data are needed to 
have a complete sample and statistically
confirm these findings.

\begin{figure}  
\begin{center}
\includegraphics[width=0.50\textwidth,height=0.35\textheight,angle=0]{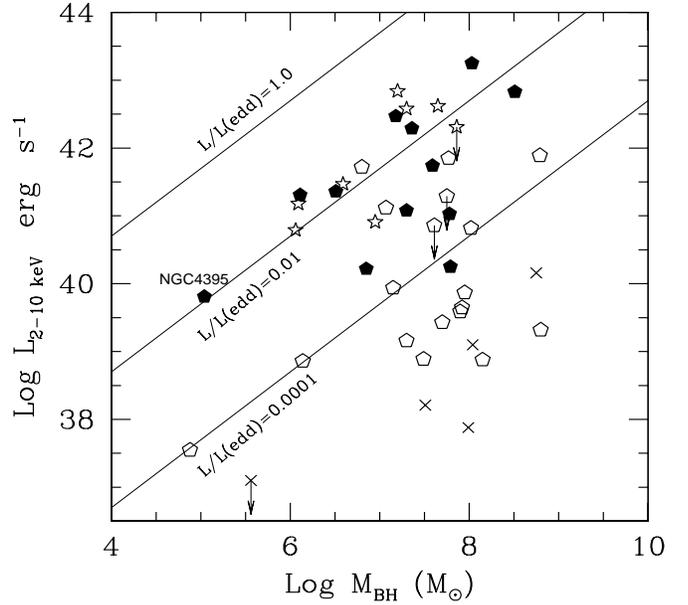}
\caption{Log of 2-10 keV luminosity versus log of black hole mass 
(symbols as in fig \ref{ox}). Compton thick candidates are plotted as stars.
The solid lines show
the L$_{2-10 keV}$ as a function of the black hole mass
for Eddington ratios of 1.0 and 0.01.}
\label{xmbh}
\end{center}
\end{figure}

In Fig.~\ref{xmbh}, Compton thick corrected X-ray luminosities 
have been plotted against black hole mass.
No correlation is found between these two parameters.
The same considerations can be applied when H$_{\alpha}$ and [OIII]
luminosities are plotted against M$_{BH}$, i.e.
no correlation with M$_{BH}$ is observed, which is
an expected result given the correlation of these 
quantities with the X-ray luminosity.
Some previous studies found
a correlation between the AGN luminosity and M$_{BH}$
(Koratkar \& Gaskell 1991, Kaspi et al. 2000),
however our result is in agreement with that found
by Pellegrini (2005) for a sample of SMBH 
in the local universe for which Chandra nuclear
luminosities were available.
Also Woo \& Urry (2002) found no correlation between 
bolometric luminosity and black hole masses.
Interestingly, neither is radio emission
correlated to black hole mass in nearby nuclei (Ho 2002).

We have over-plotted L$_{X}$ 
as a function of M$_{BH}$ for Eddington ratios of 1.0, 0.01 and 10$^{-4}$
(solid lines in Fig.~\ref{xmbh}). 
Woo \& Urry (2002) have shown that
bright local AGN normally show Eddington ratios
which span three orders of magnitude
down to L$_{Bol}$/L$_{Edd}$ $\sim$ 10$^{-3}$, while at higher redshifts
the Eddington ratios distribution is narrower, i.e.
L$_{Bol}$/L$_{Edd}$ peaks at around 1/3
with a dispersion of 0.3 dex rms, as recently
shown by Kollmeier et al. (2005) for a sample of AGN
discovered in the AGES Survey covering a redshift range of 0.3-4.
Indeed, most of our sources are radiating 
at very low Eddington ratios if compared with luminous AGN.
The low Eddington ratios observed in our sample are even lower
if the bolometric correction considered is that of LLAGN.
At such low Eddington ratios, radiatively inefficient accretion is
normally invoked as the putative mechanism for the production of the observed emission.
For example, in Merloni et al. (2003) the L$_{2-10 keV}$/L$_{Edd}$ ratio 
for an heterogeneous sample of AGN and stellar masses black holes
ranges from 1 to 10$^{-8}$.  These authors delineate a range of 
L$_{2-10 keV}$/L$_{Edd}$ ratios in which the accretion mode changes from 
a radiatively efficient to a radiatively inefficient one
below 10$^{-3}$ and above $\sim$ 0.7. According to these claims,
most of our sources would be powered by 
radiatively inefficient accretion; this issue is further discussed in Sec. 7.
Finally, note that in Fig.~\ref{xmbh} Compton thick sources
populate the upper part of the plot, at higher Eddington ratios,
while only one source has L$_{Bol}$/L$_{Edd}$ $\lesssim$ 10$^{-3}$;
this is probably a selection effect, since Compton thick sources
with an observed L$_{2-10 keV}$ $\lesssim$ 10$^{38}$ erg s$^{-1}$
would probably be undetectable in our sample.

\section{Discussion}

The X-ray analysis performed on a sample of nearby 
Seyfert galaxies has revealed the presence of a
central active nucleus in all but four sources out of 47. The intrinsic nuclear
X-ray luminosities span nearly five orders of magnitude,
down to 10$^{38}$ erg/s in the 2-10 keV energy range.
After taking into account the presence of Compton thick
objects in which the nuclear
emission is heavily reduced by absorbing material,
we have found a strong correlation between
X-ray and optical line emission luminosities. This suggests
a link between the X-ray emission and the ionization radiation
which holds also at very low luminosities.
In luminous AGN, the UV/optical radiation
has the form of a
"big blue bump" feature (Elvis et al. 1994),
commonly modelled by a geometrically thin
accretion disk working in a radiatively efficient regime
(Shakura \& Sunyaev 1973) while 
the X-ray emission is produced by Comptonization of soft
photons in a hot corona (Haardt \& Maraschi 1991). On the other hand,
it has been shown that some
LLAGN lack this "big blue bump" feature (Ho 1999).
Such observational evidence combined with the
low Eddington ratios commonly observed in LLAGN,
has lead many authors to invoke 
advection-dominated accretion flow models
(ADAF, Narayan \& Yi 1994) and their variants
to model their spectral shape.
As a matter of fact, ADAF models work in a radiatively
inefficient regime at
sub-Eddington ratios (L $<$ 0.01-0.1L$_{Edd}$)
and can reproduce the lack of UV excess observed
in the SED of LLAGN.
However, also radiatively efficient standard accretion disks
are stable at low Eddington ratios down to L $\sim$ 10$^{-6}$L$_{Edd}$
(Park \& Ostriker 2001) and can reproduce the shape
of the LLAGN SED since the temperatures of a multi-colour disk
scale with $\dot{m}$$^{1/4}$ (Ptak et al. 2004).

Several examples of sources studied in the literature have shown
that reconciling the multi-waveband observations of LLAGN
with theoretical models is a complex task. For example, 
Ptak et al. (2004) have shown that the UV/X-ray
spectrum of NGC~3998 could be reproduced equally well by
radiatively inefficient accretion flow
(RIAF) models and Comptonized thin disk models
with $\dot{m}$ $<$ 10$^{-5}$ at which a standard disk
is both stable and thick.
In the cases of M~81 and NGC~4579, two well known LLAGN
belonging also to our sample, a geometrically thin disk
extending up to $\sim$100 Schwarzchild radii was
required to account for their optical/UV spectra,
while the X-ray radiation has been explained by
an optically thin, two temperature ADAF at
smaller radii (Quataert et al. 1999). 
NGC~4258 has been considered as an ideal candidate
for ADAF model (Lasota et al. 1996), while
Yuan et al. (2002) have shown that the jet
component dominates the emission.
Based on the IR/X-ray Eddington ratio,
the X-ray variability and spectral shape,
Fiore et al. (2001b) considered NGC~4258 an AGN in a low state
or a scaled-down version of a Seyfert galaxy.

As reported in this work,
the X-ray versus optical emission line 
correlations scale with luminosity, so that
low luminosity Seyfert galaxies appear to be a scaled-down version of
classical AGN. The observed optical emission line spectra
suggest that the amount of UV radiation produced
is sufficient to ionize the NLR. 
However, if Seyfert galaxies with very low luminosities in our sample 
actually lack a "big blue bump" feature, as 
sometimes observed in LLAGN, then an additional source
of UV photons must be invoked which means that 
the AGN UV photons are not the only contributors
to the ionizing radiation, i.e. radiation produced
in circumnuclear starburst could contribute significantly
to the observed emission line fluxes.
A similar scenario has been proposed by Contini (2004)
for NGC~4579, in which both line and
continuum emission are explained by a composite
model which accounts for the presence of an 
AGN, a jet which interacts with the ISM creating shocks,
a starburst of different ages and HII regions.
Moreover, some authors have shown that
the observed emission line ratios in AGN may be
produced by both a thermal or a non-thermal optical/UV continuum 
(Keel \& Miller 1983, Martins et al. 2003).
Indeed, to constrain the spectral shape of these sources
and compare it with theoretical models, it is necessary 
to properly subtract UV/optical data for 
the Galactic emission and have 
multi-waveband observations. 

\subsection{On the effects of the incompleteness of the sample}

The Palomar sample is one of the best samples
available for the kind of study presented here. In
Ho \& Ulvestad (2001) a detailed discussion of
biases is presented and the sample
is compared with other Seyfert samples available in the literature.
In this work, only 47 out of 60 Seyfert
galaxies belonging to the original complete sample (HFS97) have X-ray data, 
and this fact certainly introduces incompleteness problems and biases
in our study. To investigate this issue, we have
looked at the H$_{\alpha}$ luminosities
of the 13 objects without X-ray data and found that they range from
38.43 $<$ Log L$_{H\alpha}$ $<$ 40.54 erg s$^{-1}$ (HFS97).
In this range of luminosities we have shown that
the X-ray vs. optical emission line correlations still
hold: therefore we should expect these sources
to have low X-ray luminosities. Of course,
given the large scatter observed at low luminosity,
a determination of the X-ray luminosity for the total sample
is important both to better calibrate the correlations
down to low luminosities and also to strengthen our conclusions
on the basis of a complete sample.
Regarding the analysis involving the black hole mass, 
the fact that no dependence between luminosity 
and M$_{BH}$ has been found, suggests that 
no significant differences in the total M$_{BH}$ distribution 
is expected, if we were able to add the M$_{BH}$ estimates
for these 13 missing sources.
Of course, the availability of M$_{BH}$ estimates 
for the complete sample of Palomar Seyfert galaxies will reduce some
bias effects, albeit, since M$_{BH}$ estimates have been obtained
by using different methods having different degrees of
uncertainty, it is intrinsically difficult to
estimate the degree of incompleteness of the M$_{BH}$ data. 

\section{Conclusions}

The X-ray analysis of a sample of 47 nearby Seyfert galaxies 
(type 1, type 2 and 'mixed Seyfert')
has allowed us to obtain nuclear 2-10 keV X-ray luminosities with minimal 
contamination by off-nuclear sources and diffuse emission.
The dichotomy often observed in the luminosity of type 1 and type 2 AGN,
is mainly due to the presence of heavy absorption in type 2 objects,
as demonstrated in C06 and confirmed here.
A sub-sample of 11 candidate Compton thick sources in our sample 
($>$ 30\% of type 2 Seyferts) has been found. Their
observed luminosities have been increased by an indicative
factor of 60 to take into account the luminosity obscured 
from our line of sight.
In the effort of further verifying the physical continuity
between our sample and bright AGN, the X-ray luminosities, 
non-corrected and corrected for Compton thick sources, 
have been correlated with the H$_{\alpha}$ and the [OIII] luminosities, 
both suspected to be absorption independent quantities 
and good tracers of the nuclear emission. X-ray luminosities
have also been correlated with M$_{BH}$. The results obtained are:

\begin{itemize}
\item Both L$_{X}$ vs. L$_{H\alpha}$ and
L$_{X}$ vs. L$_{[OIII]}$ correlations are highly significant in our sample,
indicating that the X-ray emission and the UV ionizing radiation are linked.
\item Both correlations scale with luminosity, i.e. they have similar slopes
of more powerful objects, suggesting that low luminosity Seyfert galaxies 
are powered by the same physical processes which operate 
in brighter AGN such as QSOs.
\item No correlation is found between nuclear luminosity
and M$_{BH}$ in agreement with some previous studies 
(Woo \& Urry 2002, Pellegrini et al. 2005).
\item L/L$_{Edd}$ ratios span three order of magnitudes
down to L$_{Bol}$/L$_{Edd}$ $\sim$ 10$^{-7}$, indicating
that most of our sources are accreting at very low Eddington ratios.
\end{itemize}

Overall our results suggest that Seyfert nuclei
in our sample are consistent with being a scaled-down version of 
luminous AGN, except for a small fraction of 
objects having luminosities in the 10$^{37}$ - 10$^{38}$  
erg s$^{-1}$ range and which show evidences of stellar processes 
as the underlying agent responsible for the activity. 
A forthcoming paper is in preparation in which the
multi-waveband analysis for this sample of Seyfert galaxies will be
enlarged, including the study of nuclear radio, IR and optical emission
jointly with the X-ray luminosity and M$_{BH}$ estimates, and tested
against theoretical models.

\begin{acknowledgements}
We thank Silvia Pellegrini and Andy Fabian for 
helpful suggestions. 
We thank our referee for valuable suggestions.
F.P. acknowledges support by a "Juan de la Cierva" fellowship.
Financial support for F.P., X.B. and F.J.C. was provided by the Spanish
Ministry of Education and Science, under project ESP2003-00812.
\end{acknowledgements}

\section*{APPENDIX: NOTES ON INDIVIDUAL SOURCES}\label{appex} 

In this section we report notes on individual sources.
We only discuss the X-ray data analyzed in this work; 
for those sources
with X-ray data taken from the literature
the references are reported in Table~\ref{x}.
We give a description of: (i) the nuclear X-ray morphologies 
(ii) the \xmm and \cha spectral results and (iii) results from the literature. 
Spectral best fit results are discussed only for spectra with more than 100 counts.
\\
\\
{\it NGC 1058} - In both \cha and \xmm images 
a nuclear core is absent, in agreement with the 
absence of a radio core detection 
(Ho \& Ulvestad 2001). The upper limit on the 2-10 keV luminosity reported 
for this source has been derived from the \cha observation 
assuming $\Gamma$ = 1.8, in agreement with the value 
reported by Ho et al. (2001) on the same data set. \\
{\it NGC 1275} - This object is the central galaxy of the Perseus 
cluster. The \cha image shows a compact nucleus surrounded by a
complex structure which has been 
extensively studied in recent years (Fabian et al. 2003 and 
reference therein). Since \cha public data for the nucleus suffered 
from heavy pile-up, here we use for the spectral analysis an \xmm observation
taken from the archive. The 0.5-10 keV spectrum has been extracted from a
region of 20" radius. The soft spectrum is clearly dominated by the 
diffuse thermal emission of the cluster; the 2-10
keV spectrum is described by a power-law ($\Gamma$ = 1.95$\pm$0.01) absorbed by
the Galactic column density. A prominent Fe K$\alpha$ line 
associated with the cluster is present at 
$\sim$ 6.7 keV; another at $\sim$ 6.4 keV
having an equivalent width of 272 $\pm$ 25 eV is also visible
and is likely associated with the active nucleus. 
The hard luminosity is in agreement with Churazov et al. (2003).\\
{\it NGC 2683} - A low-luminosity unresolved radio core (Irwin et al. 2000)
is coincident with the nuclear peak emission seen in the \cha
image. For this object, the 2-10 keV flux has been derived from the \cha data
assuming $\Gamma$ = 1.8 and Galactic absorption.\\
{\it NGC 3079} - See C06 for the analysis of the \xmm data. The 10" region around the
nucleus is resolved in the \cha image: a strong nuclear source is
embedded in a bubble of diffuse emission. A \cha and HST study of
the superbubble by Cecil, Bland-Hawthorn and Veilleux (2002) shows that
the optical and X-ray emissions match spatially. The radio core position is
coincident with the 2-10 keV peak. We extracted the spectrum from a
circular region of 2" in radius. The spectral parameters are not well
constrained due to the poor photon statistics, however the results
are in good agreement with the \xmm ones. We fit the data with
an absorbed power-law ($\Gamma$ = 1.71$\pm$0.5, N$_{H}$ = $\leq$
4.6$\times$ 10$^{21}$ cm$^{-2}$). The strong FeK line at 6.4 keV 
(detected at $>$ 99\% of significance with \xmm) suggests that this source 
is heavily absorbed and confirm the BeppoSAX results
which indicate that the source is Compton thick (Iyomoto et al. 2001). \\
{\it NGC 3147} - A \cha snap-shot observation for this source has
been published by Terashima \& Wilson (2003). The same
data set is analyzed here. The 0.3-10 keV image reveals a bright compact source
surrounded by very faint soft and weak diffuse emission. The 2-10 keV core is
clearly detected and restricted to a region of 2" in radius. The nuclear 
spectrum was extracted from this region but it suffers from mild pile-up.
The effect of pile-up has been corrected using the PILEUP model
in XSPEC. The spectrum is described by a power-law
($\Gamma$ = 1.88$\pm$0.15) modified by low absorption. 
These results are in agreement with those by Terashima \& Wilson (2003).
The 2-10 keV absorption corrected flux is a factor of 2.5 higher than
what measured with ASCA.\\
{\it NGC 3489} - The \cha image for this source shows faint
nuclear emission and an off-nuclear source within 5" from the nucleus.
The upper limit on the 2-10 keV flux has been derived 
assuming $\Gamma$ = 1.8 and Galactic absorption and it is in 
good agreement with the results reported in Ho et al. (2001). 
An upper limit on the radio detection is reported in 
Filho, Barthel and Ho (2002).\\
{\it NGC 3516} - This galaxy has been extensively studied in X-rays
and it is known to be variable both in flux and spectrum
(Guainazzi et al. 2001). We analyzed the zeroth-order image and spectrum of 
a 47 ks ACIS/LETGS observation (Netzer et al. 2002) and a 100 ks EPIC
observation. In the \cha image we find a bright point-like source
at the nuclear position. Soft diffuse 
emission surrounds the nucleus and it extends for 10". 
The nuclear spectrum has been
extracted from a circular region of 2" in radius, while we used 25" for the \xmm
spectrum. Both the \xmm and \cha spectra show a flat 
continuum with overimposed several narrow components of Fe K$\alpha$ 
along with a broad line (see Turner et al. 2002). Here we adopt 
the \cha results where $\Gamma$ = 1.31$\pm$0.1, N$_{H}$ $\sim$ 1$\times$
10$^{22}$ cm$^{-2}$ and the equivalent width of the iron K line 
at 6.4 keV is 96$^{+0.47}_{-0.57}$ eV. The hard spectrum obtained
is due to the fact that the source was
in a low state at the time of the observation. Despite this, 
fluxes and luminosities are in agreement 
with previous measurements.\\
{\it NGC 3608} - Only \cha data are available for this object. The image
of the nuclear region is characterized by a complex structure, where the
peak of the emission is in the nucleus but is
surrounded by off-nuclear sources. The 2-10 keV flux has been obtained
assuming a power-law with $\Gamma$ = 1.8 (fixed) and Galactic absorption.
Only an upper limit on the radio detection is available (Wrobel 1991).\\
{\it NGC 3627} - In the \cha image
some structures are visible close to the nucleus and in
the \xmm image the weak nuclear emission is comparable to
a source off-set by $\sim$ 10". 
The upper limit on the \cha 2-10 keV flux has been derived assuming a 
$\Gamma$ = 1.8 and is consistent with the result obtained by Ho et al.
(2001) on the same data set. The \xmm spectrum
is probably contaminated.
\\
{\it NGC 3982} - This source has been observed by ASCA.
We have derived the 2-10 keV flux from the 
count rate presented in Moran et al. (2001) assuming a 
$\Gamma$ = 1.8 and Galactic absorption..
An unresolved radio core has been detected (Ho \& Ulvestad 2001).\\
{\it NGC 4051} - A previous \cha HETG study for this source has
been performed by Collinge et al. (2001).  Here we use the ID 2148
observation (frame time 0.4 sec). The 2-10 keV compact nucleus of
this object is embedded in soft diffuse emission. An extraction radius
of 2" allows us to reduce the contamination in the soft band.  
The spectrum suffers from mild pile-up ($\leq$ 11\%), therefore we use the
PILEUP model in XSPEC to take into account this effect (alpha parameter = 0.29). 
We fitted the 0.5-10 keV spectrum with a soft thermal component 
with kT at 0.2 keV, a power-law plus an Fe line at 6.4 keV significant
at $>$ 99\% (EW=345$^{+49}_{-91}$ eV). The spectral index
($\Gamma$ = 1.33$^{+0.7}_{-0.03}$) and the value of the column density 
($\leq$1.4$\times$10$^{20}$ cm$^{-2}$) are in good agreement with
the spectral values obtained with \xmm except for the Fe line which is
weaker with EW = 240$\pm40$ eV (C06). Our results are in agreement
with what reported in Lamer et al (2003), i.e. that the spectral parameters for
both the continuum emission and the line
are variable. A complex radio structure is present in this source (Ho \& Ulvestad 2001).\\
{\it NGC 4235} - The ASCA hard X-ray luminosity has been obtained 
by modeling the spectrum with an absorbed power-law 
($\Gamma$ = 1.57$\pm$0.06, N$_{H}$ = 1.5$\times$10$^{21}$).
The ASCA data are taken from the HEASARC archive. 
An unresolved radio core is present in the source (Ho \& Ulvestad 2001).\\
{\it NGC 4258} - The \cha image of this object shows a prominent
nuclear emission located at the same position of the nuclear radio core (Ho
\& Ulvestad 2001). An extraction radius of 2" allows us to avoid the
contamination of an off-nuclear source positioned at only 3" from the
nucleus. This source is not resolved in the \xmm image which is
dominated by a hard point-like nucleus and unresolved diffuse
emission (C06). The \xmm and \cha spectral results are in good 
agreement with results reported by Pietsch \& Read (2002). The \cha
spectral shape is a power-law with $\Gamma$ = 1.4$\pm$0.1 and an 
intrinsic absorption of
7$\times$10$^{22}$ cm$^{-2}$. No FeK$\alpha$ line is detected in this case. 
The \xmm hard luminosity is a factor of 2 lower than the \cha
luminosity.  However, it has been shown that
both X-ray flux and spectral shape 
are highly variable in this source (Fruscione et al. 2005). 
The same is true for the iron K$\alpha$ which had been
detected in previous {\it ASCA} and {\it Beppo}SAX observations 
but which is not significantly detected both in our \cha and 
\xmm measurements.\\
{\it NGC 4388} - \cha observed this source on 2001 June (Iwasawa et
al. 2003).  The 2-10 keV image shows a bright nucleus which appears
embedded in diffuse emission in the full band image which extends for
$\sim$ 20" and appears to be correlated with the optical ionization cones. 
We selected a circular region of 2" in radius centered on the hard peak position
which agrees very well with the northern radio core supposed to be the
true nucleus (Ho \& Ulvestad 2001). The spectrum is affected by mild
pile-up. We fitted the hard strong component with an heavily absorbed
power-law having a spectral index of $\sim$ 1.7 and obtain a N$_{H}$ =
3.3$\times$10$^{23}$ cm$^{-2}$. A strong Fe K$\alpha$ line at 6.32 keV
is detected (EW $\sim$ 450 eV) while marginal is the
detection of a line at 7.0 keV. The soft component is fitted with a
power-law with $\Gamma$ = 0.4$\pm$0.2. 
Our spectral results are in good agreement with those reported in 
Iwasawa et al. (2003). The \xmm spectrum is characterized by a flat
spectral slope and the amount of absorption as well as the equivalent
width of the iron line are in good agreement with the 
\cha values. Variability in the column density has been
reported by Elvis et al.(2004).\\
{\it NGC 4472} - This is an elliptical giant galaxy. The \xmm
image reveals strong soft diffuse emission. The \cha higher angular 
resolution allows us to resolve the 25" region of \xmm in 
diffuse emission and off-nuclear sources around the optical 
nuclear position. The \cha 2-10 keV image is characterized by 
a complex structure and there is no evidence for a dominant core emission
(details on this \cha data set are given in Loewenstein et al. 2001 
and Soldatenkov, Vikhlinin, \& Pavlinsky 2003).
For this reason the hard X-ray flux and luminosity are treated as upper limits.  
This source is marginally detected in the radio band (Ho \& Ulvestad 2001). \\
{\it NGC 4579} - This object has been observed for $\sim$ 35 ks
(Eracleous et al. 2002) and $\sim$ 3 ks (Ho et al. 2001, Terashima and
Wilson 2003) by \cha. \cha images show a hard compact nucleus
surrounded by soft diffuse emission which extends for $\sim$ 40". In
both observations the nucleus is significantly piled-up ($\sim$ 11\%).
In Eracleous et al. (2002) the spectral fitting
has been corrected from the pile-up effect by using the simulator-based
forward-fitting tool LYNX, developed by the ACIS instrument team.
The data are described by a simple power-law with
no absorption in excess to the Galactic value ($\Gamma$ =
1.88$\pm$0.03). 
Here we derive the 2-10 keV fluxes and luminosities
from the 3 ks observation and find that they are in good agreement
with Eracleous et al (2002) and Ho et al. (2001) results. 
A radio core is detected (Ho \& Ulvestad 2001). \\
{\it NGC 5194} - This galaxy, also called M 51, has been observed by \cha four times.
A 15 ks observation has been analyzed by Terashima \& Wilson (2001);
here we consider the 26 ks observation. The \cha image shows a
complex nuclear region characterized by extended features and
off-nuclear sources. A bright nucleus is seen in the optical position
coincident with the radio core position (Ho \& Ulvestad 2001) and 
it appears to be compact in the 2-10 keV image. The nuclear spectrum
has been extracted from a region of 2" in radius. The soft 
emission has been modeled with a thermal plasma with kT = 0.32$\pm$0.4 keV
while the hard component with a power-law having photon index 
fixed to $\Gamma$ = 2; the strong iron line detected at $\sim$ 6.4 keV has an
equivalent width greater than 2.6 keV, which is an 
indication of the Compton thick nature of this source.
This has also been confirmed by a BeppoSAX observation of M 51 which has
shown that the nucleus is absorbed by a column density of
5.6$\times$10$^{24}$ cm$^{-2}$ (Fukazawa et al. 2001).\\
{\it NGC 5548} - This source has been observed by \cha using the 
LETG (Kaastra et al. 2002) and the HETG (Yaqoob et al. 2001) instruments. 
The zeroth-order observations are affected by heavy pile-up. We analyzed
the 92 ks \xmm observation available from the archive.
The \cha and \xmm images reveal the presence of a point-like
bright nucleus. We extracted the \xmm nuclear spectrum from a
region of 25" in radius. A power-law with $\Gamma$ = 1.69$\pm$0.01
modified by galactic absorption gives a good fit of the spectrum. An
FeK line is detected (EW = 67$^{+37}_{-7}$ eV) plus a prominent soft excess.
Our best fit is consistent with the work on the same data set 
by Pounds et al. (2003).\\
{\it NGC 6482} - This source has been observed by \cha for $\sim$
20 ks. The nuclear emission is very bright, diffuse and typical of an
elliptical galaxy. 
We extract the spectrum from a
circular region of 2" in radius centered in the optical position which
is coincident with the 2-10 keV peak. The
spectrum is dominated by a soft thermal component (kT = 0.8$\pm$0.5 keV),
while the hard component has been modelled with an absorbed 
($\leq$ 1.45$\times$10$^{21}$ cm$^{-2}$) power-law having
the photon index fixed to 1.8. This source has not been detected
at 8.4 GHz (Filho, Barthel, \& Ho 2002).\\
{\it NGC 6503} - This galaxy has been observed by \cha twice
(exposures of 2 ks and 13 ks). Here we analyze the longer exposure
observation. The nuclear position is determined by the comparison
of the \cha image with the HST image. The nucleus has been detected,
although it is very weak. Four off-nuclear sources have also been detected. 
Few counts are extracted from a region of 2" in radius from which 
flux and luminosity have been derived assuming $\Gamma$ fixed to 1.8. 
Emission has not been detected in the radio band (Filho, Barthel and Ho 2000).\\
{\it NGC 7479} - We report the analysis of the 50 ks \xmm
observation of this source (Iwasawa et al. in preparation). From a
comparison between the EPIC images and the DSS image we clearly
detect the nuclear source which appears to be fainter than the few
off-nuclear sources associated with the asymmetrical spiral arms. The
nuclear spectrum has been extracted from a circular region of 20" in
order to avoid contamination. However, the 2-10 keV image reveals the
possible presence of unresolved sources within 20" around the nucleus.
To model the continuum we fixed the photon index of the power-law to
1.8 and left the column density free to vary 
(N$_{H}$ = 5.8$\times$10$^{23}$ cm$^{-2}$). We also fit the soft component 
using a thermal model (kT = 2.0$\pm$0.5 keV). 
The radio core is unresolved (Ho \& Ulvestad 2001).\\

\end{document}